\documentclass[pdflatex,sn-nature]{sn-jnl}

\usepackage{graphicx}%
\usepackage{multirow}%
\usepackage{amsmath,amssymb,amsfonts}%
\usepackage{amsthm}%
\usepackage{mathrsfs}%
\usepackage[title]{appendix}%
\usepackage{xcolor}%
\usepackage{textcomp}%
\usepackage{manyfoot}%
\usepackage{booktabs}%
\usepackage{algorithm}%
\usepackage{algorithmicx}%
\usepackage{algpseudocode}%
\usepackage{listings}%
\usepackage[normalem]{ulem}
\usepackage{placeins}
\usepackage{geometry}
\usepackage{setspace}
\usepackage{geometry}

\newcommand{\sima}{${\sim}$}
\newcommand{\rein}{$R_\mathrm{Ein}$}
\newcommand{\reff}{$R_\mathrm{e}$}
\newcommand{\Mein}{$M_\mathrm{Ein}$}
\newcommand{\Mstar}{$M_\mathrm{\ast}$}
\newcommand{\Meinstar}{$M_\mathrm{\ast,R_{Ein}}$}

\newcommand{\fDM}{$f_\mathrm{DM,R_{Ein}}$}
\newcommand{\fDMdep}{$f_{\mathrm{DM,3D},r_\mathrm{e}}$}
\newcommand{\sersic}{S{\'e}rsic}
\newcommand{\alphaimf}{$\alpha_\mathrm{IMF}$}
\newcommand{\pr}{{\rm P}}

\begin{document}

\title[Milky-Way-like stars in a galaxy core 8 billion years ago revealed by gravitational lensing]{Milky-Way-like stars in a galaxy core 8 billion years ago revealed by gravitational lensing}


\author[1,*]{Quirino D'Amato}
\author[1]{Filippo Mannucci}
\author[2]{Alessandro Sonnenfeld}
\author[3,4,1]{Martina Scialpi}
\author[5]{James W. Nightingale}
\author[6]{Cristiana Spingola}
\author[1,4]{Stefano Zibetti}
\author[4,1]{Alessandro Marconi}
\author[7]{Piero Rosati}
\author[4,1]{Cosimo Marconcini}
\author[1]{Guido Agapito}
\author[1]{Anna Gallazzi}
\author[4]{Enrico Di Teodoro}
\author[9,17]{Gloria Andreuzzi}
\author[1,8]{Francesco Belfiore}
\author[1]{Elena Bertola}
\author[4,1]{Caterina Bracci}
\author[10]{Stefano Carniani}
\author[4,1]{Elisa Cataldi}
\author[1]{Avinanda Chakraborty}
\author[4,1]{Matteo Ceci}
\author[11]{Claudia Cicone}
\author[12]{Anna Ciurlo}
\author[1]{Giovanni Cresci}
\author[13]{Alessandra De Rosa}
\author[14]{Elisa Di Carlo}
\author[1]{Anna Feltre}
\author[4,1]{Michele Ginolfi}
\author[1,4]{Isabella Lamperti}
\author[4,1]{Bianca Moreschini}
\author[1]{Emanuele Nardini}
\author[15]{Michele Perna}
\author[14]{Elisa Portaluri}
\author[11]{Khatun Rubinur}
\author[16]{Paolo Saracco}
\author[16]{Paola Severgnini}
\author[17]{Vincenzo Testa}
\author[18]{Giulia Tozzi}
\author[10]{Giacomo Venturi}
\author[3,4,1]{Lorenzo Ulivi}
\author[19]{Cristian Vignali}
\author[1]{Maria Vittoria Zanchettin}
\author[20]{Antonio Pepe}

\affil[*]{Corresponding author. E-mail: quirino.damato@inaf.it}
\affil[]{Affiliations are reported after the references}
\affil[]{}
\affil[]{The peer-reviewed version of the paper is published by Nature Astronomy, available at \url{https://www.nature.com/articles/s41550-026-02819-4}}
\affil[]{DOI number: 10.1038/s41550-026-02819-4}


\abstract{The assembly of stellar‑dominated cores in elliptical galaxies is key to understand how cosmic structures evolved. Gravitational lensing offers unique insights into the nature of their stars. We report the discovery of the smallest known quadruply-lensed quasar (radius$\sim$0.2"), whose lensing galaxy at redshift 1.055 (5.5 billion years after the Big Bang) features a lensing mass of only $\sim2 \times 10^{10}~\mathrm{M_\odot}$. Bayesian analysis, based on the system's exceptional properties and standard scaling relations, allowed to sample the central galactic initial mass function with unmatched accuracy and in a previously uncharted regime in terms of mass and redshift. We found it consistent with the Milky Way one, while excluding bottom-heavy functions. This suggests that the core either grew slowly or underwent early disruptive events altering its stellar build-up, in contrast with the classical view that bulges form rapidly and remain unchanged by later interactions.}

\newgeometry{top=2cm}
\maketitle
\restoregeometry

Despite decades of study, several fundamental questions remain about the internal structure and evolution of elliptical galaxies, also known as early-type galaxies (ETGs). A widely discussed scenario, predicted by cosmological simulations \cite{oser_2010} and supported by observational results \cite{zibetti_2020}, is the ``two-phase'' evolution: a rapid ($<$1 Gyr) dissipative collapse in the first $\sim3$ billion years of the Universe (at redshifts $z~{>}$ 2), followed by later assembly through gas-poor mergers with satellite galaxies. Such evolutionary processes are expected to affect the number of stars formed per stellar mass bin and unit volume, known as the initial mass function (IMF), since the accreted object may have in general a different IMF from the preexisting population \cite{van_dokkum_2010}. The classical two-phase model predicts that the center of ETGs is dominated by low-mass stellar populations (described by a bottom-heavy Salpeter IMF; \cite{salpeter_1955}), that remain mostly untouched until the present days \cite{oser_2010}. In contrast, the Milky Way is well described by the so-called bottom-light Chabrier IMF \cite{chabrier_2003}, possibly as a result of a steady and secular evolutionary path. The universality of the IMF (i.e., whether it is the same across different environments and epochs) remains an open question, due to the large uncertainties on the observational parameters and the difficulty of reconciling measurements derived with different methods (\cite{cappellari_2012,shajib_2024}). 

Galaxy-scale strong gravitational lensing is a powerful tool for constraining the IMF of ETGs \cite{treu_2010b,shajib_2024}. Lensing occurs when a foreground galaxy magnifies a background source, producing multiple images or arcs. This effect enables precise measurement of the total enclosed mass within the Einstein radius (\rein), which for quadruply lensed systems approximates the circle defined by the images. Lensing has often been combined with stellar dynamics to separate dark matter (DM) from stars, the dominant mass components in ETGs. Lensing–dynamics (LD) studies, largely based on the Sloan Lens ACS (SLACS) sample, typically favored a Salpeter IMF for massive (stellar mass $M_\ast \geq 10^{11}~\mathrm{M_\odot}$) low-redshift ETGs \cite{treu_2010a, auger_2010b, sonnenfeld_2015, shajib_2021}. However, the lack of spatially-resolved data led to rely on simplifying assumptions, like spherical symmetry and orbital isotropy \cite{auger_2010b, sonnenfeld_2015}, leading to an incorrect density slope determination \cite{barnabe_2011}. Moreover, SLACS selection effects favor high velocity dispersion systems, further skewing IMF estimates toward heavier values \cite{sonnenfeld_2024}. 

Lensing-only studies avoid these dynamical biases. Their main uncertainty lies in the assumed DM profile, though this also affects LD and dynamics-only approaches \cite{cappellari_2013, tortora_2018, forrest_2022}. However, DM can be modeled with hydrodynamical simulations (accounting for baryonic feedback, \cite{schaller_2015}) or neglected entirely to yield stringent IMF constraints. Importantly, when \rein~ lies within the stellar half-light effective radius \reff, lensing primarily probes baryon-dominated regions where DM degeneracy is minimized. The SINFONI Nearby Elliptical Lens Locator Survey (SNELLS, \cite{smith_2015}) provided key results using 3 massive ($\gtrsim 2 \times10^{11}~\mathrm{M_\odot}$) local ($z<0.05$) lensing ETGs, featuring $R=r_\mathrm{ein}/r_\mathrm{eff} <1$. The sample's analysis showed that lensing-only methods imply a Kroupa-like, bottom-light IMF, while inclusion of dynamics yields systematically heavier IMFs \cite{newman_2017}. The authors argued that lensing masses are more robust, primarily because even when DM is excluded the inferred IMF remains lighter than Salpeter. The origin of discrepancies between lensing and dynamical estimates remains unclear. Other important source of uncertainties can contribute to the observed bias, such as different stellar population synthesis (SPS) modeling and different physical scales probed by lensing and  dynamics. A comprehensive discussion of possible sources of bias in lensing IMF studies is reported in Appendix \ref{subsec:uncer_comp}.

While past and current works focused on massive ETGs, the IMF of lower-mass galaxies ($\log(\mathrm{M_\ast}/\mathrm{M_\odot})~\lesssim 10.6$)~remains essentially unconstrained due to their rarity as lenses and observable uncertainties. LD and dynamical studies of massive ETGs highlighted an IMF evolution with both the dynamical mass and redshift \cite{cappellari_2013,posacki_2015,mendel_2020,forrest_2022}, whose extrapolations possibly suggest lighter IMF at lower masses, but direct constraints are still lacking. At high-$z\gtrsim0.5$, the IMF of such a population is unknown, since no lensing ETGs of this class have been detected so far.

\begin{figure}
	\centering
	\includegraphics[width=0.7\textwidth]{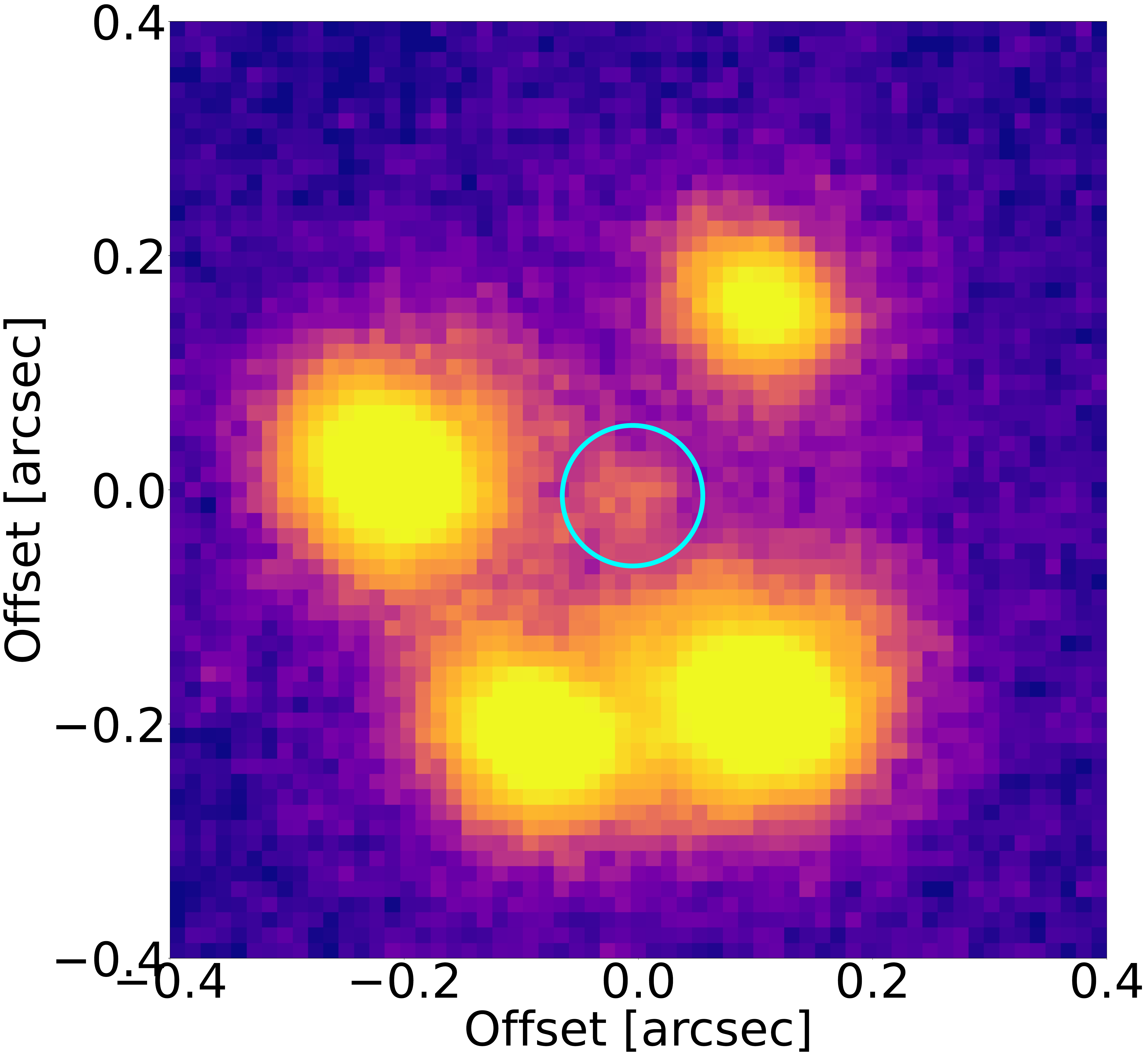} 
	\caption{\textit{Ks}-band image of the object obtained with the ERIS NIX imager. The lensing galaxy is visible at the center of the quadruply-lensed quasar configuration, marked by the cyan circle. 
        }
	\label{fig:image} 
\end{figure}

\begin{figure} 
	\centering
	\includegraphics[width=\textwidth]{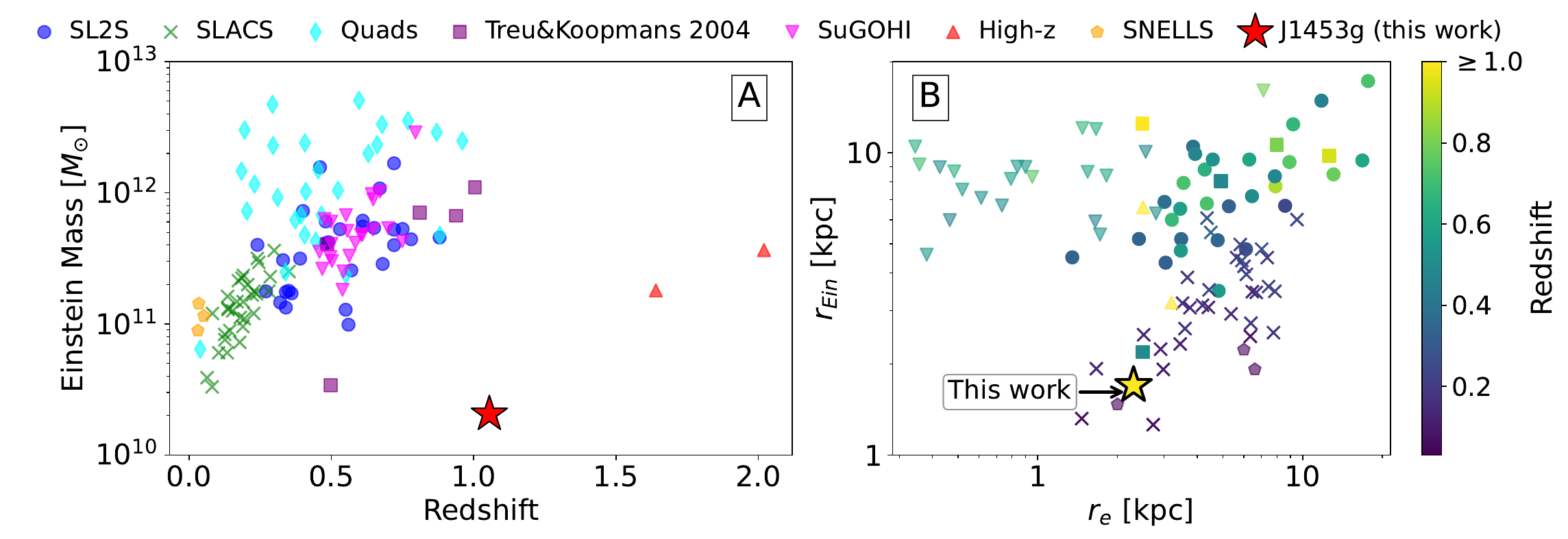} 
	\caption{(A) \Mein~ as a function of redshift of different lensing samples: the 
    Sloan Lens ACS (SLACS, green crosses, \cite{treu_2010a, auger_2010b, sonnenfeld_2015}), the Strong Lensing Legacy Survey (SL2S, blue circles, \cite{sonnenfeld_2013, sonnenfeld_2015}), LD galaxies from \cite{TK_2004} (purple squares), Survey of Gravitationally-lensed Objects in Hyper SuprimeCam (SuGOHI, magenta down-triangles, \cite{sonnenfeld_2019}), SINFONI Nearby Elliptical Lens Locator Survey (SNELLS, orange pentagons, \cite{smith_2015}), and high-$z$ lensing ETGs (red up-triangles, \cite{wong_2014,mercier_2024}). We also include, only for comparison purposes, several quadruple lenses collected by the Gravitationally Lensed Quasar Database and for which the source and lens redshifts are known (GLQ, cyan diamonds, \cite{GLQ}). The target of this work (J1453g) is marked by the red star. (B) \rein~ as a function of \reff~ for the same samples of panel A, except for the GLQ database, color-coded by the lens redshift. 
        }
	\label{fig:comparison} 
\end{figure}

In this work, we present the discovery of the first low-mass ETG lens. The lensing system (J1453+0520) is composed by a quasar at $z=2.82$ lensed by a galaxy (hereafter J1453g) at $z=1.055$ ($\sim$5.5 Gyr after the Big Bang). The background quasar lies within the lens tangential caustic curve, producing a quadruple image in the plane of the sky (see Figure \ref{fig:image}). The system features the smallest angular separation among quadruply lensed quasars discovered so far, and it is the second smallest gravitational lens ever detected \cite{york_2005}. In addition, it features the smallest Einstein mass \Mein~(i.e., the total mass within \rein)~among all known lenses across all redshifts and for which the IMF has been investigated, pushing this measurement in a completely unexplored parameter space (see Figure \ref{fig:comparison}, left panel, for a comparison with other ETGs lensing samples). The quadruply-lensed quasar configuration of J1453+0529 enables a precise estimate of \rein~and, in turn, of \Mein. Importantly, J1453g is among the few high-$z$ lensing systems having $R<1$, probing the star-dominated central region (see Figure \ref{fig:comparison}, right panel). Thanks to these peculiarities, we measured  
the IMF in the galaxy inner region with exceptional accuracy, addressing the possible sources of uncertainties discussed above, finding it to be consistent with the Chabrier IMF. We reject a Salpeter IMF with high significance, supported by formal Bayesian analysis (equal to the method widely used in LD and lensing studies, \cite{treu_2010a,sonnenfeld_2015,sonnenfeld_2025}), conservative assumption and hydrodynamical prescriptions. In contrast, the classical two-phase model predicts that the galaxy center is dominated by lower-mass stellar populations, that remain mostly untouched until the present days \cite{oser_2010}. 

\section{Results}
\subsection{Observations and galaxy properties}
J1453+0529 was first identified as a multiple point-like object using the Gaia Multi Peak (GMP) technique \cite{mannucci_2022}, a method that identifies multiple peaks in the light profiles of previously selected quasars in the \textit{Gaia} archive \cite{krone-martins_2024}, with separations down to \sima0.15$''$ \cite{mannucci_2023}. Subsequent adaptive optic (AO) spectroscopy and imaging follow-up performed with the ERIS instrument \cite{pearson_2016,davies_2023}, at the Very Large Telescope (VLT), revealed the quadruple lensing nature of the system, with a background quasar at $z_\mathrm{s}=2.82\pm 0.01$. The ERIS NIX \textit{Ks}-band image in Figure \ref{fig:image} shows the four images of the background quasar and the central lens galaxy (see also Subsection \ref{subsec:light_fit}).

Spatially-unresolved spectrography performed with the DOLORES instrument, at the Telescopio Nazionale Galileo (TNG), revealed the presence of deep,  narrow absorption lines due to inter-stellar material superimposed to the quasar spectrum from a single, galaxy-scale, absorbing system, that we identify as the lensing galaxy. We fitted the spectrum and measured a lens redshift of $z_\mathrm{l}=1.055\pm 0.001$ (Subsection \ref{subsec:TNG_spec}).

We performed the lensing and light profile modeling using the state-of-art {\sc{PyAutoLens}} package \cite{nightingale_2018,nightingale_2021b}. We performed a Bayesian posterior probability fitting  of the multiple quasar image positions, modeling the total lensing mass with a singular isothermal ellipsoid (SIE). This model assumes a power-law mass profile with a fixed slope $\alpha=2$, and is ubiquitously used to describe galactic scale lensing mass profiles (\cite{shajib_2021, tan_2024}, see also Appendix \ref{subsec:DM_caveats}). We tested several fitting prescriptions (Subsection \ref{subsec:lens_fit}), consistently deriving a robust estimate of the most important parameter, \rein, which only depends on the well-constrained quadruply-lensed quasar positions.

As for the image modeling, we simultaneously modeled the combined light profile of the four quasar images and the lens, using a recursive Bayesian fitting (see Section \ref{subsec:light_fit}). We obtained the best fit when an elliptical S{\'e}rsic light profile \cite{sersic_1963} with index $n=4$ (typical of ETGs) is used to model the lens. In addition, the inferred \reff~ for the lens is in excellent agreement with that expected for elliptical galaxies with similar mass and redshift \cite{van_der_wel_2014}. Different assumptions on the stellar profile do not change our conclusions (see Appendix \ref{subsec:stellar_size_unc}). We estimated the total observed stellar mass $\mathrm{M_\ast}$ of the lens using its measured $K$-band magnitude $m_{Ks, \mathrm{lens}}$ and assuming either a Salpeter and a Chabrier IMF. We used the mass-to-light ratio (M/L) conversion presented by \cite{longhetti_2009}, specifically tuned for $K$-band observations of ETGs, that assumes a galaxy formation redshift $z_\mathrm{f} =4$ (corresponding to a galaxy age of ${\sim}$4 Gyr; see Subsection \ref{subsec:gal_age} for age-dependent stellar mass and age constraints). The fitting parameters relevant to the following analysis are reported in Table \ref{tab:fitting}. A complete list of the best-fit parameters is presented in Subsection \ref{subsec:light_fit}.

\begin{table} 
	\centering
	\caption{Main galaxy properties from light and lens fitting.
				From left to right: \textit{Ks}-band magnitude, half-light radius, total stellar mass assuming a Chabrier and a Salpeter IMF, the Einstein radius and the Einstein mass. For each parameter, the uncertainties are reported at the 99.7\% confidence Bayesian fit sample distribution. }
	\label{tab:fitting} 
	
	\begin{tabular}{cccccc} 
		\\
		\hline
		$m_{Ks, \mathrm{lens}}$ & \reff & $M_{\ast,\mathrm{obs}}^\mathrm{Chab}$ & $M_{\ast,\mathrm{obs}}^\mathrm{Salp}$ & $r_\mathrm{Ein}$ & \Mein \\
                                & (kpc) & $M_\odot$               & $M_\odot$   & (kpc) &   $M_\odot$ \\
 		\hline
        $19.4_{-0.1}^{+0.1}$ &  $2.3_{-0.02}^{+0.02}$ & $3.1_{-0.2}^{+0.4} \times 10^{10}$ & $5.6_{-0.4}^{+0.6} \times 10^{10}$ & $1.71_{-0.15}^{+0.17}$ & $2.05_{-0.3}^{+0.4} \times 10^{10}$ \\
		\hline 
	\end{tabular}
\end{table}

\subsection{The nature of the stars in the galaxy core}
\label{subsec:nature_stars}
\begin{figure} 
	\centering
	\includegraphics[width=\textwidth]{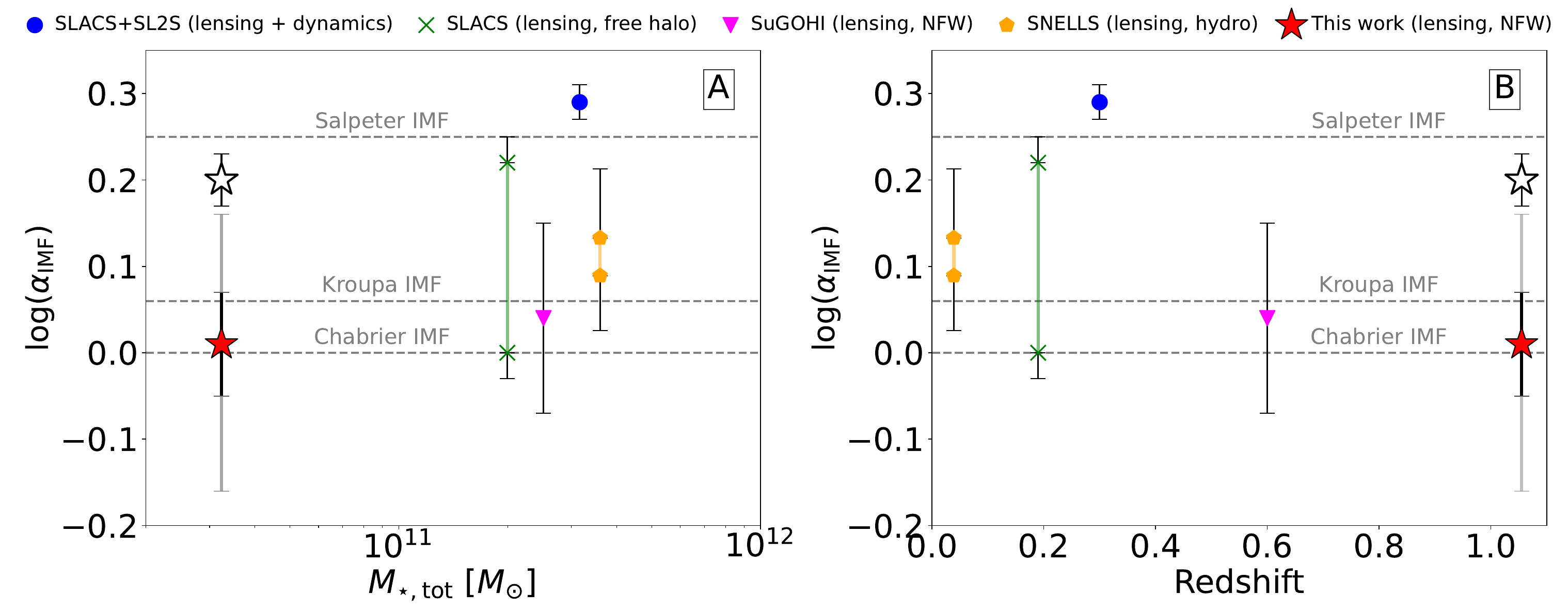} 
	\caption{(A) We show the \alphaimf~ parameter inferred from different surveys with different methods, as a function of average \Mstar. The range marked by the solid green line for the SLACS sample stands for different level of halo contraction, where no contraction corresponds to the highest point (DM halo described by a NFW profile). The range marked by the solid orange line for the SNELLS sample indicates different assumption on the galaxy age. For the different samples, error bars are at the 68\% significant level. For our object (red star), black and gray error bars are at the 68\% and 99.7\% level, respectively. The upper limit case, assuming no DM, is marked by the black open star. (B) Same as panel A, \alphaimf~ is plotted against the sample average redshifts.
        }
	\label{fig:alphaimf} 
\end{figure}

A key method used to compare different IMFs, widely adopted in lensing and LD studies over the past two decades, is the calculation of the so-called ``IMF mismatch parameter'' \cite{treu_2010b, shajib_2024}, defined as:
\begin{equation}\label{eq:alphaIMF}
\alpha_\mathrm{IMF} = \frac{M_\ast^\mathrm{True}}{M_\ast^\mathrm{Chab}}.
\end{equation}
$M_\ast^\mathrm{Chab}$ is the stellar mass that would be measured by fitting perfect photometric data with no errors assuming a Chabrier IMF, while $M_\ast^\mathrm{True}$ is the unknown true stellar mass of the galaxy. This parameter quantifies how much stellar mass a galaxy truly has, compared to what one would infer assuming a given IMF. In our case (where the reference IMF is Chabrier), if \alphaimf~ $= 1$, the galaxy’s actual stellar mass matches the Chabrier-based estimate. Values higher than 1 indicate more low-mass stars (i.e., a bottom-heavier IMF, like Salpeter), while values below 1 suggest a bottom-lighter IMF. To estimate \alphaimf, we combined the lensing and photometric measurements. We constructed a Bayesian framework that relates all the observed quantities to a set of standard scaling relations, including empirical connections between the galaxy size, the stellar mass, and DM halo properties (Subection \ref{subsec:IMF_infer}). For the DM halo we adopt a Navarro–Frenk–White (NFW, \cite{navarro_1996}) profile, as widely adopted in LD and lensing studies \cite{TK_2004, treu_2010a, sonnenfeld_2015, shajib_2021, shajib_2024, tan_2024}. We also include a prior probability in the Bayesian analysis that takes in account the selection function for the GMP method \cite{mannucci_2022}, to correct for selection biases. From our Bayesian analysis, we derive $\log$(\alphaimf) $= 0.01 ^{+0.15}_{-0.17}$ at the 99.7\% confidence level (see Figure \ref{fig:alphaimf}), strongly favoring a Chabrier IMF for our object and rejecting at $>3\sigma$ a Salpeter IMF (corresponding to $\log$(\alphaimf)$= 0.25$). We report in  Figure \ref{fig:alphaimf} a comparison with other lenses' samples. 

\section{Discussion}
\subsection{Robustness of the results}
\label{sec:robustness}

Our measurement avoids key limitations by using a system with a small $R$, precise \rein~ measurement offered by the lensing configuration, and by building the likelihood framework on state-of-art cosmological relations. As mentioned in the introduction, the main sources of uncertainties for lensing-only studies are the assumption of a DM profile and, as for all the techniques, the SPS modeling. As for the DM profile, first of all we notice that lensing-only reanalysis of the SLACS sample indicates that maximum halo contractions (corresponding to a steeper-than-NFW dark matter profile) favor a Chabrier IMF, whereas minimum contractions (i.e., NFW profile) yield a Salpeter IMF \cite{sonnenfeld_2025}. In this respect, our Bayesian analysis is performed assuming an NFW profile and robustly finds $\alpha_\mathrm{IMF}$ consistent with a Chabrier IMF; thus, our conclusion is not affected by the IMF-halo contraction degeneracy. Secondly, we can derive a stringent constraint under the extreme assumption of complete absence of DM, following the test performed by \cite{newman_2017} for the SNELLS survey. We notice that this scenario is strongly disfavored by both hydrodynamical and N-body simulation prescriptions (see below and Subsection \ref{subsec:IMF_infer}) and by the best-fit lensing modeling that we obtain (Subsection \ref{subsec:lens_fit}), which corroborates the well-established ``bulge–halo conspiracy'' scenario where a non-negligible amount of DM is present (see Appendix \ref{subsec:DM_caveats}). In addition, large‐sample analyses of low‐$z$ lensing ETGs find DM profiles consistent with NFW or slightly steeper forms \cite{shajib_2021,tan_2024,sheu_2024}. We repeated the Bayesian analysis in absence of DM and assuming that the stellar mass within \rein~ is equal to the total mass \Mein~ (see Subsection \ref{subsec:IMF_infer_noDM}), finding a stringent upper limit to $\log$(\alphaimf) $= 0.20 ^{+0.03}_{-0.03}$ at the 68\% confidence level. As expected, the IMF normalization required to reproduce the observed total mass is considerably higher, but, given it as extreme upper limit, still significantly below the Salpeter $\log$(\alphaimf) = 0.25. In other words, even under this unphysical assumption, the Salpeter IMF provides a stellar mass higher than the observed total mass.

We can also tackle the same problem from the opposite point of view with a simple calculation, that is the estimate of the projected dark matter fraction \fDM~ $=M_\mathrm{DM,r_{Ein}}/M_\mathrm{tot,r_{Ein}}$, where $M_\mathrm{tot,r_{Ein}} = M_\mathrm{Ein}$ is fixed by lensing at \rein, for the two different IMF assumptions. In fact, once the stellar mass within $r_{\mathrm{Ein}}$ (\Meinstar) is set by the IMF and light profile, the projected DM mass is simply \Mein $-$ \Meinstar, which can yield \fDM $<0$ if \Meinstar $>$ \Mein, that is if the stellar mass for a given IMF exceeds the total mass. We found $f_\mathrm{DM} = 0.36^{+0.08}_{-0.08}\,(\mathrm{stat})\,^{+0.04}_{-0.08}\,(\mathrm{syst})$ for a Chabrier IMF and
$f_\mathrm{DM} = -0.14^{+0.14}_{-0.15}\,(\mathrm{stat})\,^{+0.12}_{-0.08}\,(\mathrm{syst})$ for a Salpeter IMF (See Subsection \ref{subsec:DM_frac}). In other words, a Salpeter IMF would imply a stellar mass larger than the total mass measured by lensing inside the Einstein radius. Even at its largest error bound, a Salpeter IMF yields $f_\mathrm{DM}\sim0.12$, implying an extremely expanded DM halo, while hydrodynamical simulation consistently predict halo contraction \cite{cautun_2020,schaller_2015}. In addition, hydrodynamical simulations predict a three-dimensional DM fraction $\gtrsim40\%$ within \reff~ for $M_\ast\sim10^{10-11}\,M_\odot$ over $z=0 -1$ \cite{lovell_2018,mukherjee_2022}. Assuming a spherical NFW halo and Chabrier IMF, the deprojected DM fraction within \reff~ is \fDMdep ~ $ = 0.27~ (^{+0.03}_{-0.06})_\mathrm{stat}~(^{+0.06}_{-0.06})_\mathrm{sist}$, rising to $\sim0.37$ under maximal contraction. For a Salpeter IMF, even with maximal contraction, we obtain at maximum \fDMdep~ \sima~ 10\%. Thus, a Salpeter IMF is in much greater tension with simulation predictions. Finally, we notice that our result agrees with expectations obtained from extrapolating the relation between \alphaimf~ and SIE velocity dispersion inferred by LD studies of high-mass low-$z$ ETGs (see Appendix \ref{subsec:LD_comp}).

As for the SPS modeling, the main uncertainty in the observed band arises from the unknown galaxy's age. Assuming an age-dependent M/L \cite{longhetti_2009}, we reject at the 99.7\% confidence level a bottom-heavy IMF if the galaxy is at least $\sim$3.3 Gyr old (formation redshift $z_\mathrm{f} \sim2.9$), while we reject it at the 90\% confidence level if the galaxy is $\sim$3 Gyr old (formation redshift $z_\mathrm{f} \sim2.5$). Although we cannot directly measure the age of the galaxy with our current data, our Bayesian analysis provides not only the stellar IMF normalization ($\alpha_\mathrm{IMF}$) but also an estimate of the galaxy's DM halo mass and size (Subsection \ref{subsec:IMF_infer}). Using these properties, we can infer the halo formation redshift (which is suggested to be coeval with the beginning of the star formation \cite{donnari_2019}), by comparing with predictions from DM and hydrodynamical simulations. Based on the halo's accretion history, we estimated that it assembled about 3.9 billion years before the galaxy's observed redshift (see Subsection \ref{subsec:gal_age}), in excellent agreement with the $z_\mathrm{f} =4$ assumption made by \cite{longhetti_2009}. Even accounting for the scatter of these predictions, we found that the halo must have formed at least 3.5 billion years before observation. In other words, if the galaxy was significantly younger, it would imply a dark matter halo too massive for its age, something that current simulations and empirical scaling relations strongly disfavor.

In conclusion, with our measurement at \rein/\reff~ $\approx 0.74$, coupled with the precise measurement of \rein~ offered by the quadruply-lensed configuration of the system, we report the most robust constraint of a galaxy's bulge IMF obtained so far for a single object, based on the many arguments presented above. With a \Mstar~$\sim 10^{10}~ \mathrm{M_\odot}$ and $z\sim1$, the galaxy is also the first of its class in terms of mass and redshift, opening a new window on an uncharted parameter space of galactic IMF-constraining lensing studies. 

\subsection{Implications for the evolution of elliptical galaxies}
\label{sec:gal_ev}
In the two-phase galaxy formation scenario, the central regions of ETGs are thought to originate from an early, intense starburst phase at redshifts $z \gtrsim 2$, producing compact systems often referred to as ``red nuggets'' \cite{oldham_2017}. The star formation in this early phase occurs on short timescales ($< 1$ Gyr) in dense, turbulent, and high-pressure environments. Most models predict that the IMF is bottom-heavy for stars formed via this rapid channel \cite{chabrier_2023, shajib_2024}. Over time, these galaxies grow in size primarily through minor mergers and secular star formation, building up their outskirts with stars formed under a comparatively bottom-light IMF, as observed in local Universe galaxy disks \cite{chabrier_2003}. This classical two-phase scenario is supported by the discovery of the ``relic galaxies'' \cite{spiniello_2021}, that are very massive (\Mstar~$\gtrsim  10^{12}~ \mathrm{M_\odot}$) galaxies in the local Universe evolved passively across cosmic time and considered the local counterpart of the red nuggets. Simulations suggest that, regardless of stellar mass, the bulges of elliptical galaxies form rapidly and through in-situ (merger-free) star formation; for \Mstar~$\sim 10^{10}~ \mathrm{M_\odot}$ galaxies, in particular, this process is predicted to be dominant at all radii \cite{pulsoni_2021}. However, in principle, the inner galaxy region can be also diluted or destroyed by later accretion events, potentially producing an evolution of the central IMF towards a bottom-lighter IMF, as a result of a slower and steady star formation history \cite{shajib_2024}. So far, there are no observational evidence that disruptive events are able to significantly change the initial mass function and chemical composition of galaxy central regions.

The observed bottom-light IMF already in place at $z=1$ could be in principle explained if the galaxy has been building gradually via time-extended secular processes since its formation epoch. However, both theoretical and observational results of local and high-$z$ ETGs (based on scaling relations and metal abundances studies, \cite{longhetti_2009,gallazzi_2006}) indicate that most of the stellar mass is built in a short star formation timescales ($< 1$ Gyr). The subsequent phase of passive evolution, characterized by dry, minor mergers, can cause a growth in size up to a factor of $\sim$5, but only a few percent in stellar mass \cite{oser_2010, naab_2009}. If J1453g initially formed through a rapid ($\lesssim 1$ Gyr), early starburst, producing a bottom-heavy IMF core, our finding suggests that the galaxy core experienced one disruptive event -- or several minor accretion episodes -- within a span of few Gyr after the initial mass building process, able to drastically change its stellar IMF observed at $z\sim1$. 

These findings challenge the classical two-phase scenarios in which the galaxy bulge forms rapidly \textit{and} remains structurally and chemically unchanged between $z\sim3$ and $z\sim1$, for \Mstar~$\sim 10^{10}~ \mathrm{M_\odot}$~ ETGs. Finally, we note that it is tempting to connect our results to the SNELLS survey at $z \sim 0$ (reporting similar bottom-light IMF at similar $R$ values), possibly suggesting that the IMF remains relatively stable from $z \sim 1$ to the present day. However, we point out that the SNELLS galaxies probe a high stellar mass regime (\Mstar~$\gtrsim 10^{11}~ \mathrm{M_\odot}$), and thus cannot be considered direct descendants of our system.

\section{Methods}\label{sec:methods}

\subsection{Target selection}
\label{subsec:tgt_sel}

The quasar J1453+0529 (RA=14:53:06.95,  Dec=+05:29:28.06) was selected using the Gaia Multi Peak (GMP) technique \cite{mannucci_2022}. This method looks for the presence of multiple peaks in the 1-D light profile of the \textit{Gaia} sources associated to previously classified quasars. \cite{mannucci_2023} showed that this technique is sensitive to objects with separations between 0.15", which is the limit set by the point spread function (PSF) of the instrument, and $\sim0.7"$, above which most of the sources are well resolved and appear as multiple entries in the \textit{Gaia} archive. The GMP method has proven to be a very effective way to select physical pairs of active galactic nuclei (AGN) and multiply imaged, strongly lensed quasars at sub-arcsec separations \cite{ciurlo_2023, mannucci_2023,scialpi_2024}. This is especially true at $z > 0.3$, where the host galaxy does not contribute significantly to the \textit{Gaia} light profile \cite{gaia_collab_2023}.

J1453+0529 was classified as a quasar with G=19.41 at $z=2.79$~ by the LAMOST survey \cite{yao_2019, jin_2023}. It was observed as a part of a project aimed at discovering multiple and lensed AGN at sub-arcsec separation using the ERIS GTO time. The selected targets satisfy the GMP selection \cite{mannucci_2022}, are at high galactic latitudes, and are far from large, local galaxies such as the Magellanic Clouds. All targets have known redshifts to put one of the main emission lines in the ERIS wavelength bands, and have suitable natural guide stars (NGSs) to drive the AO system.

\subsection{ERIS resolved spectroscopy and images}
\label{subsec:resolv_spec}

ERIS \cite{davies_2023} is composed by two different instruments, the near-IR camera NIX \cite{pearson_2016}, and the integral-field unit (IFU), near-IR spectrograph SPIFFIER \cite{george_2016}.
J1453+0529 was originally observed with SPIFFIER on April 14th, 2023 (program ID = 111.24QJ.001) to obtain resolved spectroscopy of the system at the diffraction limit of the telescope. The presence of four sources at small separations was immediately evident in the spectra, revealing the lensed nature of the system. The lens galaxy was not detected either in emission or in absorption, due to the low signal-to-noise ratio (S/N) of the spectrum. The NIX camera was used the following night to obtain deep imaging of the system, in the $H$ (exposure time: 900 s) and $Ks$ bands (exposure time: 1500 s), and reveal the presence of the lens galaxy. A relatively large scale of 26 mas/px, was used to maximize the surface brightness sensitivity, at the expense of a sub-optimal sampling of the PSF, whose diffraction-limited size would be ${\sim}$ 40 mas in the $H$ band and ${\sim}$55 mas in the $Ks$ band. These preliminary images both revealed a clear detection of the central lensing galaxies, but the poor PSF sampling and the lack of a usable PSF in the field of view prevented us from performing accurate image analysis and PSF reconstruction necessary to measure the galaxy properties. We subsequently reobserved the system with NIX on February 1st, 2024 (program ID: 112.25GT.001), in $Ks$ band, using a pixel scale of 13 mas/px for optimal PSF sampling, and a longer exposure time (1800 s). The image, used in this work for both light and lensing analysis (Figure \ref{fig:image}), has been obtained using the ESO NIX pipeline version 1.5.3 with default parameters and recipe workflow.

\subsection{TNG integrated spectrum}
\label{subsec:TNG_spec}

 Typically, in gravitational lensing systems, the redshift of the lens can be determined by analyzing absorption features imprinted on the background quasar's spectrum as its light passes through the lensing galaxy. However, due to the low S/N of the integrated spectrum obtained from the LAMOST survey \cite{yao_2019, jin_2023}, we were unable to measure the redshift of the lensing galaxy. For this reason, we observed J1453+0529 on May 06, 2024, using DOLORES, a low-resolution spectrograph and CCD camera mounted on the Telescopio Nazionale Galileo (TNG), as part of the \textit{``Giovani Astronomi e Telescopio Nazionale Galileo''} program. 
 The total integration time was 45 minutes, utilizing a $1''$ long-slit and the LR-B grism (spectral resolution of $R \simeq 585$), which covers a wavelength range of 4085--8430 $\AA$. The target was observed at the parallactic angle, with an average seeing of $0.9''$ during the exposure. The star FEIGE34 was observed within the same night to flux-calibrate the data. We reduced and flux-calibrated the data using the {\sc{PypeIt}} data reduction pipeline~\cite{prochaska20}, a Python package designed for semi-automated reduction of astronomical spectroscopic data.

\subsection{Spectrum analysis and redshift of the lens}
\label{subsec:spec_anal}

\begin{figure} 
	\centering
	\includegraphics[width=0.9\textwidth]{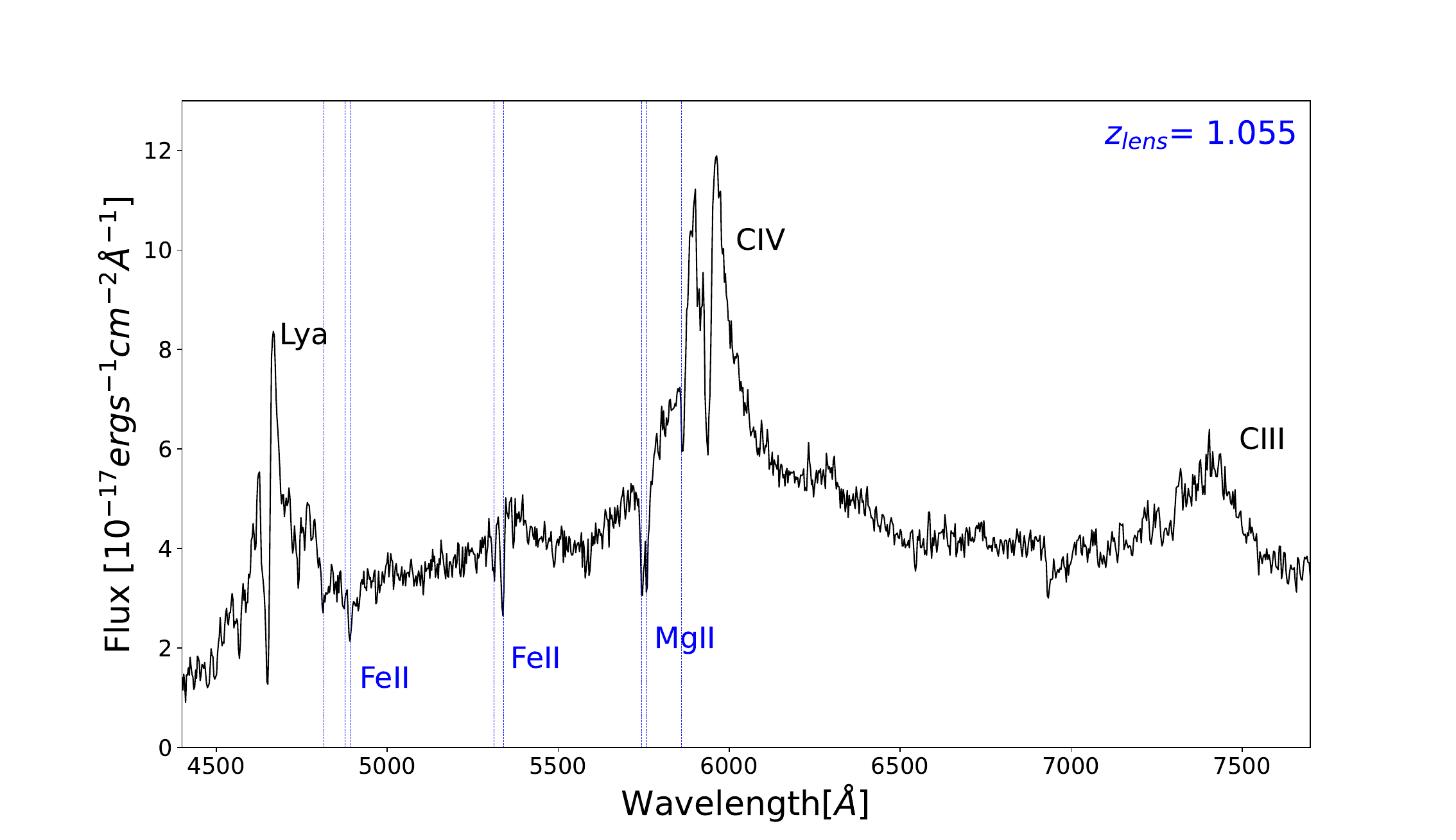} 
	\caption{TNG spectrum. The black curve is the total spectrum, while blue vertical lines indicate the absorption lines of the lensing galaxy at $z=1.055$.
        }
	\label{fig:tng_spec}
\end{figure}

The extracted total spatially unresolved spectrum obtained with TNG is shown in Figure \ref{fig:tng_spec}, where several broad and narrow emission line from the quasar are detected with high S/N. 
We fitted the spectrum with an empirically-derived spectral energy distribution (SED) for luminous Type 1 AGN \cite{Temple_2021}, covering the rest-frame ultraviolet to near-infrared band, including dust attenuation and a boot strap analysis to estimate the uncertainties on the fitted physical quantities. From the fit we obtain $z_\mathrm{s}=2.84 \pm 0.05$, in good agreement with the value of $2.79$ derived from the low S/N LAMOST spectrum. The prominent lines in the quasar TNG spectrum are Ly$\alpha$, CIV and CII, which are known to be possibly Doppler-shifted due to, e.g., the presence of outflows, affecting the redshift uncertainties. Then, using the same spectral fitting procedure as above, we fitted the combined four lensed spectra of the quasar from the ERIS/SPIFFIER data, in which the H$\gamma$ and H$\delta$ lines are detected, providing a more robust redshift estimate of $z_\mathrm{s}=2.82 \pm 0.01$, in agreement at $3\sigma$ with the other measurements.
In the TNG total spectrum, we clearly observe strong narrow absorption lines from a single absorbing system at a unique redshift (shown in blue in Figure \ref{fig:tng_spec}). We fitted the absorption lines with an interactive procedure based on the IDL MPFIT package \cite{markwardt_2009}, which utilizes the Levenberg-Marquardt technique to solve the least-squares problem. In our case we detected both Fe~II and Mg~II multiple features at the same redshift, while CaII H, K and G absorptions at $z\sim$1 falls outside the spectrum (Figure \ref{fig:tng_spec}). We estimated the redshift error from the standard deviation of the fitted absorption lines, measuring $z_\mathrm{l}=1.055\pm 0.001$. 

In general, whenever more than one absorbers are found along the line of sight between the observer and the quasar, it is impossible just from the integrated spectrum to assess which one is the lensing galaxy (e.g., \cite{smette_1995}). Nevertheless, whenever only one absorber is observed, it is always associated with the lensing galaxy when the multiple image separation is $\lesssim 1-2$ arcsec (e.g., \cite{surdej_1988, hall_2002, inada_2005,oguri_2008, more_2016, agnello_2018a, lemon_2020}). In our case, the lens position is known because it is detected in the image and because the lens reconstruction, which takes as input only the quasar multiple image positions, provides a lens center position consistent with the peak emission of the detected galaxy. Thus, given the lens position (equally distant from each of the multiple quasar image) and the system scale, the light path must go through the inner and gas-richer part of the lensing galaxy. Therefore, in order to not be the lens, there should be another single strong absorber at $z=1$ within the system area (${\sim}$0.4" diameter) from the system center causing the absorption, while the lens itself should not produce any absorption even in the galaxy central part. The equivalent widths (EWs) of the detected absorption features implies that they are associated to a galaxy-scale system; however, there are not other objects detected in the ${\sim}$25" field of view (FOV) of the $Ks$ band NIX image. For example, for the [Mg~II]$\lambda2796\mathrm{\AA}$ we measure a EW of ${\sim}$2.2 $\mathrm{\AA}$, typical of galaxy masses of $\sim\log(M_\ast/\mathrm{M_\odot})=10.5$ (as the mass that we measured for our object) in a large range of redshift ($0.4\leq z \leq1$; \cite{das_2025}).
We do not find any other significant absorber at other redshifts. The other absorption lines visible in the spectrum are atmospheric residuals or Narrow Absorption Lines (NALs) at the quasar redshift, which are ionized ultraviolet lines associated with outflowing gas  \cite{hamann2012narrowuvabsorptionline, Misawa_2007}.

\subsection{Lens light profile and subtraction}
\label{subsec:light_fit}

We performed the lens modeling and source reconstruction exploiting the {\sc{PyAutoLens}} package for strong gravitational lensing \cite{nightingale_2018,nightingale_2021b}. {\sc{PyAutoLens}} is primarily designed to simultaneously model the lens galaxy mass profile and source light profile in lens reconstruction and we use it for lensing modeling. {\sc{PyAutoLens}} is based on {\sc{PyAutoGalaxy}} \cite{nightingale_2023}, a package that uses non-linear search algorithms for the analysis and fitting of galaxy light profiles, and is then perfectly suited also for image fitting when several components with different shapes are present. 

An accurate PSF model is needed to both fit the four point-like AGN and to convolve the lens extended model used to fit the lens. However, our object is isolated in the sky and there are no point-like sources in the ${\sim}$25"$\times$25" FOV to be used to reconstruct the PSF. Moreover, no sufficient telemetry parameters have been recorded to efficiently model the PSF with AO PSF modeling algorithms (e.g., {{\sc{TIPTOP}}; \cite{neichel_2020}). In addition, since the ERIS instrument guides on the NGS and not on the target, the shape of the PSF is direction-dependent in the FOV, elongated towards the NGS. Nevertheless, we can obtain a good estimate of the PSF and an accurate deconvolution of the quasar and the lens galaxy by exploiting the presence of the four point-like images of the AGN. Given the small angular scale separation between the components, their PSF are virtually identical and can be used for PSF modeling. 

We adopted a recursive approach to simultaneously fit the four AGN and the lensing galaxy, using the AGN fit result to convolve the lens galaxy profiles, until a good global fit was obtained. We proceeded as follows: we initially fitted only the four AGN to obtain an initial guess on the PSF model to use in the first iteration to convolve the lens galaxy model. Since the AO PSFs usually display complex shapes, we reproduce the PSF with different combinations of elliptical Moffat and Gaussian functions. In particular, we fitted the PSFs with a single Moffat, single Gaussian, two Moffat, two Gaussian, and a combination of a Moffat and a Gaussian functions. The functional form of the 2D Moffat is \cite{moffat_1969}:
\begin{equation}\label{eq:moffat}
I_m(r) = A_m \left[ 1 + \left( \frac{r}{R_m} \right)^2 \right]^{-\beta},
\end{equation}
where $A_m$ is the amplitude, $R_m$ is the width parameter and $\beta$ is a constant determining the concentration of the function. For an elliptical Moffat function $R = \alpha/\sqrt{q}$ where $q$ is the axis ratio and $\alpha$ is the circular width parameter. The function is defined on a elliptical coordinates system where $r$ is the radial distance from its center, and where the position angle $\Phi$ (defined counter-clockwise from the positive $x$-axis) and the axis ratio $q$ can be expressed in terms of elliptical components ($e_x,e_y$), by the following relations:
\begin{equation}\label{eq:ellcomp}
\begin{array}{ccc}
e_x &=& f \times \cos(2\phi) \\
e_y &=& f \times \sin(2\phi),
\end{array}
\end{equation}
where $f = (1-q)/(1+q)$. As for the Gaussian function the adopted form is:
\begin{equation}\label{eq:gauss}
I_g(r) = A_g \exp \left[ -\frac{1}{2} \left( \frac{r}{R_g} \right)^2 \right],
\end{equation}
where $A_g$ is the amplitude, $R_g = \sigma/\sqrt{q}$ is the width parameter, and $\sigma$ is the Gaussian circular width. The function is defined on elliptical coordinate system as described in Eq. \ref{eq:ellcomp}.

 For the lens we adopted an elliptical S{\'e}rsic light profile:
\begin{equation}\label{eq:sersic}
I_s(r) = A_s \exp \left\{ -b_n \left[ \left( \frac{r}{r_e} \right)^{1/n} -1 \right]  \right\},
\end{equation}
where $A_s$ is the amplitude, $r_e$ is the effective radius, $n$ is the S{\'e}rsic index determining the concentration of the function, and $b_n$ is a constant defined such that $r_e$ encloses half of the total light (For $n> 0.36$, $b_n$ can be approximated to a fourth grade polynomial function of $n$ \cite{ciotti_1999}). The S{\'e}rsic function is defined on elliptical coordinate system as described in Eq. \ref{eq:ellcomp}. 

We recursively fitted the image with the four AGN and lens models with {\sc{PyAutoLens}} via the {\sc{Nautilus}} non-linear search algorithm \cite{nautilus}, updating the convolving PSF at each step with the best-fit PSF of the previous step, until the fit did not provide significant improvements of the residuals map and Bayesian evidence. We repeated the same procedure for the Gaussian PSF.
For the three combinations of Moffat and Gaussian functions described above, where each AGN is fitted with two components, we additionally constrained the center position and the amplitude ratio of the two components of the PSF to be the same for each AGN. For each PSF model, we explored varying several parameters of the S{\'e}rsic profile. To facilitate converge toward a correct solution, we fixed the model lens center at the position of its peak in the image. 
As for the S{\'e}rsic index, we found that it is poorly constrained when left as a free parameter in the fit, so we explored two fixed values: $n=1$ (typical for spiral galaxies), and $n=4$ (typical for elliptical galaxies). The best fit in terms of residuals and Bayesian evidence is obtained for $n=4$. In particular, lens residuals for $n=1$ are of the order of $\sim30-40\%$ of the peak (depending on the PSF modeling), while for $n=4$ the residuals are $\sim15\%$ (that is within the root mean square of the image pixels).

Lensing objects are generally found to be elliptical galaxies (e.g., \cite{faure_2008,oguri_2010}), and our galaxy clearly displays a compact emission indicating the presence of a bulge. In addition, the $r_e$ resulting from the $n=1$ is of the order of $\sim$0.13" -- 0.16" in combination with all the PSF models, corresponding to $\sim$1.05 -- 1.3 kpc at the source redshift. For spiral galaxies with a stellar mass and redshift similar to our target, typical effective radii are found to be much larger ($\sim$5 -- 7 kpc; \cite{van_der_wel_2014}). In contrast, for the best fitting PSF models (see below) in combination with a lens S{\'e}rsic profile with fixed $n=4$ we obtain $r_e\sim 0.21 - 0.28$", corresponding to $\sim$1.7 -- 2.3 kpc, consistent with the expected effective radius of elliptical galaxies with similar mass and redshift \cite{van_der_wel_2014}. In addition, we note that the fit with fixed $n=4$ clearly returns the best residuals and Bayesian evidence, regardless of the PSF modeling. Finally, we note that the EW measured for the [Mg~II]$\lambda2796\mathrm{\AA}$ (${\sim}$2.2 $\mathrm{\AA}$) is compatible with the ISM properties expected for quiescent galaxy with the mass of our object sampled in the central kpc \cite{das_2025}.

With regards to the PSF modeling, single (either Moffat or Gaussian) fitting functions do not provide acceptable solutions, featuring high ($\gtrsim$20\%) residual fraction in the brightest AGN image. The fitting is significantly improved when modeling the PSF with two components. In general, the double Gaussian fitting shows larger residuals than the model composed by either two Moffat or Moffat + Gaussian functions (brightest AGN residual $\sim$13\%). These last two composed models feature excellent and comparable AGN subtraction (brightest AGN residual $\sim$8 -- 10\%) and the best Bayesian evidence. As the final best fit we selected the combination between the Moffat and Gaussian functions, due to the lowest AGN residuals and slightly lower Bayesian information criterion. However, all the fits with two-components PSF models show 1-$\sigma$ compatible effective radius, axis ratio and position angle of the lens S{\'e}rsic profile. In addition, they show very similar lens-to-total emission fraction. In particular, with the two Moffat and Moffat + Gaussian PSF modeling the lens emission accounts for $\sim$9.1\% and $\sim$8.6\% of the total emission, respectively.
This shows that the results on the lens galaxy are relatively stable with respect to the choice of the function used to reproduce the PSF. The relevant S{\'e}rsic profile parameters of the lens fitting are reported in Table \ref{tab:lens_fit}. For each parameter, the uncertainties are reported at the 99.7\% confidence Bayesian fit sample distribution. The data, model and residuals are shown in Figure \ref{fig:light_fitting}.

\begin{table} %
	\centering
	\caption{Photometric lens properties. 
				From left to right: $K_s$ band magnitude; effective radius; axis ratio; position angle; S{\'e}rsic index (fixed); stellar mass assuming a Chabrier IMF; stellar mass assuming a Salpeter IMF.}
	\label{tab:lens_fit} 

\begin{tabular}{ccccccc}
\hline
$m_{Ks, \mathrm{lens}}$ & $r_e$ & $q_{\rm l}$ & $\Phi_{\rm l}$ & $n$ & $M_\ast$ (Chabrier IMF) & $M_\ast$ (Salpeter IMF) \\
 & kpc &  & deg &  & $M_\odot$ & $M_\odot$ \\
 \hline
$19.4_{-0.1}^{+0.1}$ & $2.3_{-0.02}^{+0.02}$ & $0.72_{-0.01}^{+0.01}$ & $-2.6_{-1.4}^{+1.4}$ & 4 & $3.1_{-0.2}^{+0.4} \times 10^{10}$ & $5.6_{-0.4}^{+0.6} \times 10^{10}$ \\
\hline
\end{tabular}
\end{table}

  \begin{figure} 
	\centering
	\includegraphics[width=0.9\textwidth]{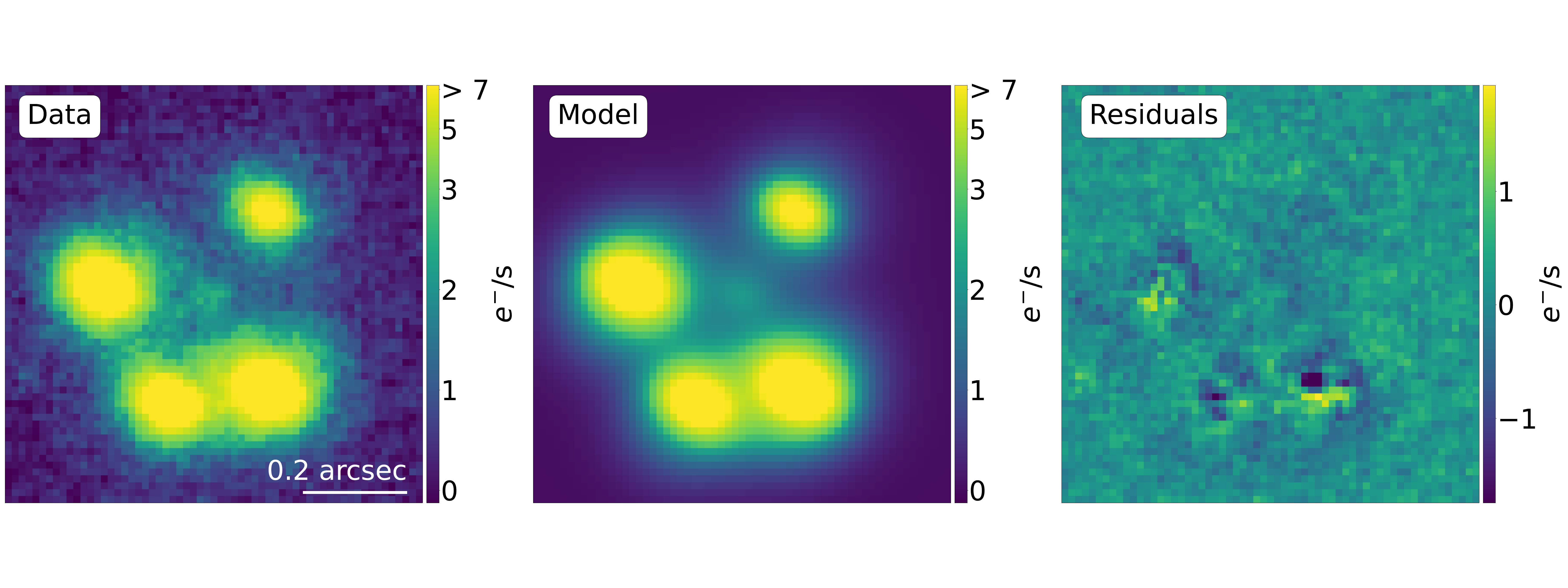} 
	\caption{From left to right: Data, model and residuals of the simultaneous fitting of multiple AGN point-like and lens S{\'e}rsic profiles.
        }
	\label{fig:light_fitting} 
\end{figure}

\subsection{Lens photometry and stellar mass}
\label{subsec:lens_phot}

We first measured the background-subtracted aperture magnitude of the entire system. The pipeline provides a zero-point and its error in the image header. However, this zero-point is based on a calibration table used by the pipeline and it is not derived from a standard star observations. Using this zero-point, within an aperture of 1.2" radius (which encompasses all the object emission), we found a magnitude $m_{Ks, \mathrm{tot}} = 16.7$. 
To check the zero-point accuracy provided by the ERIS pipeline and assess the error, we downloaded and reduced NIX observations (with the same observational setup of the target) of the two temporally closest photometric stars, taken approximately two days and one month before the target, finding a difference of -0.1 and +0.1 with respect to the target observation zero-point, respectively. Using these two zero-points for the two stars, we measured an aperture photometry in perfect agreement ($\sim 1 \sigma$) with their 2MASS catalogue magnitudes, confirming that the zero-points are correct. Thus, in order to encompass all the uncertainties due to the aperture photometry, source fitting and zero-point calibration, we conservatively assume a fiducial error of 0.1 on the target measured magnitude. We note that this value is $\sim3.5\times$ larger than the zero-point error provided by the pipeline. We consider this error as a systematic uncertainty in the physical quantities derivation along the work. 

We derived the magnitude of the lens galaxy, which, according to our model, accounts for 8.6\% of the total emission in $Ks$-band. By assigning the total system's measured magnitude to the total model emission, we derived a lens magnitude $m_{Ks, \mathrm{lens}} = 19.4$. 
From the lens galaxy magnitude $m_{Ks, \mathrm{lens}}$ we estimated the stellar mass $\mathrm{M_\ast}$ following the approach of \cite{longhetti_2009}:
\begin{equation}\label{eq:longhetti}
\log(\mathrm{M_\ast}) = \log(M_\ast/L) - 0.4 (m_{Ks, \mathrm{lens}}-d_\mathrm{mod}-k_\mathrm{corr})+1.364,
\end{equation}
where $\log(M_\ast/L)$ depends on the source redshift, age, observed wavelength, SPS model, metallicity and assumed IMF. $d_\mathrm{mod}$ is the distance modulus, while $k_\mathrm{corr}$ is the $k$-correction that depends on the source redshift, age, observed wavelength and metallicity. We note that the infrared bands are especially suited to measure single-band stellar masses of galaxy, as at $z\sim$1 sample the peak of the stellar emission, and the expected fluxes, are nearly independent on the assumed metallicity and age models compared to the optical bands \cite{longhetti_2009}. For example, the M/L ratio variation for a metallicity between solar $\mathrm{Z_\odot}$ and $0.2\times \mathrm{Z_\odot}$ is only $\sim$10\%. Similar variations are obtained when varying the exponentially declining star-formation rate time-scale (always assumed $<$1 Gyr) and stellar models. For these reasons, we fixed the metallicity to solar and adopted the conversion values reported by \cite{longhetti_2009} for the popular spectro-photometric code originally proposed by \cite{BC03}. This approach yields the lower relation scatter when compared with multi-band mass measurement (0.04 dex, \cite{longhetti_2009}). For dependence of M/L on the galaxy age, \cite{longhetti_2009} found that in $K$-band it is relatively weak, but still significant for our purpose: for a 4 Gyr old galaxy, a 1 Gyr uncertainty on its age translates to $\sim$0.2 dex variation in M/L, which is of the order of the difference between the masses derived with the Chabrier and Salpeter IMFs. The galaxy formation redshift assumed by \cite{longhetti_2009} to calculate the M/L conversion factor that we used in this work is $z_\mathrm{f} =4$, meaning that the age of J1453g at $z=1$ is $\sim4$ Gyr. We discussed the constraints of the galaxy age in Subsections \ref{subsec:nature_stars} and \ref{subsec:gal_age}.

\subsection{Lens modeling}
\label{subsec:lens_fit}

We performed the lens modeling exploiting the {\sc{PyAutoLens}} package. For point-like lensed objects and a given lensing mass profile model, the ray-tracing fitting calculates the positions of the multiple images appearing in the image-plane and evaluates the posterior probability of the model parameters. We decided to use only the multiple image positions as input for the fitting, since their flux ratios indicate that the quasar emission can possibly be affected by micro-lensing. For the AGN image positions we used the center of the four PSFs derived by fitting the light profiles. The fit returns sub-pixel error for the positions. However, we conservatively assume one pixel error on the multiple images positions.
As for the mass model, initially we adopted a general power-law elliptical profile:
\begin{equation}\label{eq:pow_mass_prof}
\rho(r) \propto r_\mathrm{Ein}/r^{\alpha}
\end{equation}
where $r$ is the elliptical radius and $\alpha$ is the slope. A particular case of the adopted profile is the SIE, where $\alpha=2$. In the image plane grid this profile can be described by a set of parameters: the coordinates of the ellipse center $\Delta x_{\rm lens}$ and $\Delta y_{\rm lens}$, the coordinates of the quasar center $\Delta x_{\rm source}$ and $\Delta y_{\rm source}$ (both defined as the RA and DEC offsets with respect to the lens light profile center, respectively), the Einstein radius, the slope, and the elliptical components as described in Eq. \ref{eq:ellcomp}. 

The initial fitting was performed by letting all the parameters vary in a broad, uniformly distributed, range of values, also encompassing clearly non-physical ranges for the Einstein radius and the source positions. This fit returned a well-constrained set of parameters apart from the elliptical components (i.e., the parameters related to the axis ratio and position angle). These parameters have two distinct peaks in the posterior probability distribution, corresponding to two ellipses of comparable size and axis ratio, but rotated by ${\sim}90^\circ$ from each other. This is because, in lens reconstruction constrained by only the positions of the point-like multiple images, the same model can be rotated by $90^\circ$ and provide very similar results, unless the mass profile is significantly stretched in a preferred direction. In addition, all the fits return $\Delta x_{\rm lens} = 0$ and $\Delta y_{\rm lens} = 0$ (i.e. centered on the peak of the lens emission) and $\alpha =2$ within 1$\sigma$, in agreement with results of previous studies about massive lensing galaxies up to $z\sim 1$ (see Appendix \ref{subsec:DM_caveats}). To better constrain the model parameters (and in particular $r_\mathrm{Ein}$), we then decided to fix the lens position to the peak emission center and to adopt an SIE model. We also restricted the value of \rein~ and of the source position to a more reasonable range, keeping a uniform prior distribution. In particular, \rein~ was free to vary by out to $\pm 0.05$" from the radius of the circle connecting the four multiple images. 
The source position was fixed to lie in a square of ${\sim}7\times7$ pixels centered at the lens peak emission. The position angle and the axis ratio can be still any between $0^\circ$ and $180^\circ$ and between 0 and 1, respectively. This final fitting removed the degeneracy on the position angle and axis ratio. It also provided the highest Bayesian Evidence. We also tried to add a shear component to the fit, but the resulting shear parameters were largely unconstrained, and such model had a lower Bayesian Evidence. 

All the models and fits we tested provide a very consistent measurement of the Einstein radius. The quadruple image nature of our objects, indeed, allows for a very precise measurement of \rein, relying on four independent constraints. This is particularly important as this is the only parameter of the lens reconstruction that we actually use in the subsequent analysis and discussion, since the Einstein mass only depends on $r_\mathrm{Ein}$, $z_{s}$, $z_l$ and the assumed cosmology (see also Appendix \ref{subsec:DM_caveats}). The maximum likelihood model critical (red) and caustics (blue) curves are displayed in Figure \ref{fig:lens_recon} superimposed to the $K_s$-band image. The reconstructed source position is marked by the red cross. The multiple AGN positions are also shown (yellow crosses). From the lens Einstein radius we calculated the Einstein mass as:
\begin{equation}\label{eq:einmass}
M_\mathrm{Ein} = \pi r_\mathrm{Ein}^2 \Sigma_\mathrm{cr}
\end{equation}
where $\Sigma_\mathrm{cr}$ is the critical surface density. Given the speed of light $c$, the gravitational constant $G$, the source angular distance $D_\mathrm{s}$, the lens angular distance $D_\mathrm{l}$, and the angular distance between the source and the lens $D_\mathrm{ls}$, the critical surface density is equal to:
\begin{equation}\label{eq:critsurfden}
\Sigma_\mathrm{cr}=\frac{c^2 D_\mathrm{s}}{4\pi G D_\mathrm{ls} D_\mathrm{l}}
\end{equation}
The best fit results are reported in Table \ref{tab:lens_fit_pyautolens}. The reported values are the median of the Bayesian fit sample distribution, while the uncertainties are at 99.7\% confidence. 

\begin{table}
	\centering
	\caption{Results of the SIE lens reconstruction.
				From top to bottom: Einstein radius; Einstein mass; mass profile axis ratio; mass profile position angle; $x$-offset of the source from the lens peak emission (fixed); $y$-offset of the source from the lens peak emission.}
	\label{tab:lens_fit_pyautolens}

\begin{tabular}{cccccccc}
\hline
$r_\mathrm{Ein}$ & $M_\mathrm{Ein}$ & $q_{\rm m}$ & $\Phi_{\rm m}$ & $\Delta x_{\rm lens}$ & $\Delta y_{\rm lens}$ & $\Delta x_{\rm source}$ & $\Delta y_{\rm source}$ \\
kpc & $M_\odot$ &  & deg & arcsec & arcsec & arcsec & arcsec \\
$1.71_{-0.15}^{+0.17}$ & $2.05_{-0.3}^{+0.4} \times 10^{10}$ & $0.73_{-0.20}^{+0.17}$ & $60.4_{-5.3}^{+5.3}$ & $0$ & $0$ & $-0.012_{-0.008}^{+0.013}$ & $-0.013_{-0.009}^{+0.013}$ \\
\hline
\end{tabular}
\end{table}

  \begin{figure}
	\centering
	\includegraphics[width=0.5\textwidth]{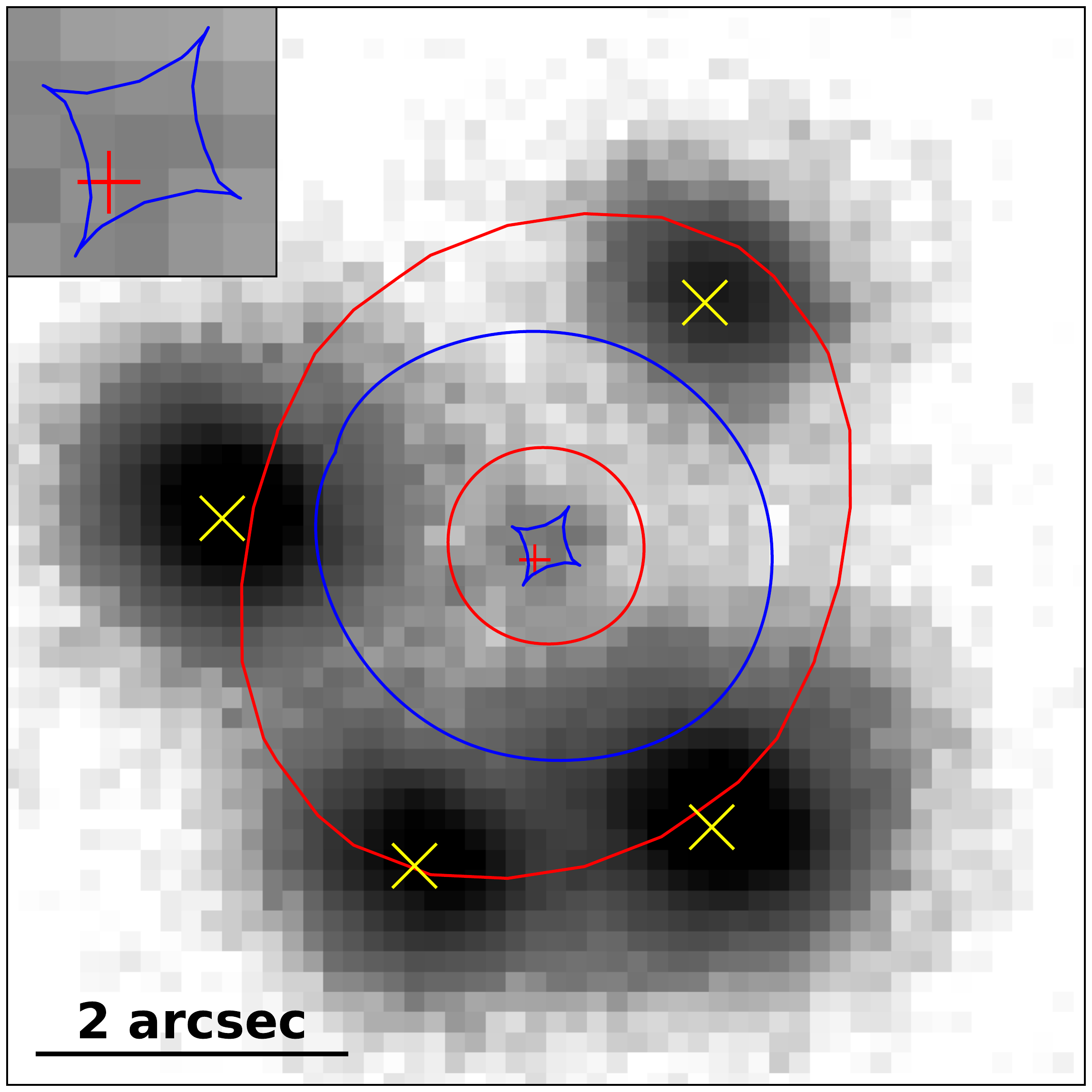} 
	\caption{Critical curves and caustics curves superimposed to the data image (solid red and blue lines, respectively). The four AGN image positions derived from the light profile fitting and used for the lens modeling are also displayed with yellow crosses. The inset shows the central part within the around the tangential caustic curve centered on the lens peak, where the reconstructed source position is marked by the red cross. The reported positions and curves are from the maximum likelihood model.
        }
	\label{fig:lens_recon} 
\end{figure}

\subsection{Stellar initial mass function mismatch parameter: Bayesian inference}
\label{subsec:IMF_infer}

Following the statistical approach of \cite{sonnenfeld_2018}, we estimated the true stellar $M_\ast^\mathrm{True}$, provided the measured \reff, \rein, $z_l$ and $z_s$ and $M_{\ast,\mathrm{obs}}^{\mathrm{Chab}}$, and allowing for the contribution of a dark matter halo to the total mass. 
To this aim, we assumed prior probabilities provided by a halo mass function, a stellar-to-halo mass relation and a stellar mass-size relation. We also assumed a fixed density profile for the baryonic and dark matter, coupled with a DM mass-concentration relation. 

For the stellar distribution, we assumed a \sersic~ profile with $n=4$, equal to the one used to fit the light profile.
For the dark matter distribution, we assumed an NFW profile:
\begin{equation}\label{eq:nfw}
\rho_\mathrm{DM}(r) \propto \frac{1}{\left(\frac{r}{r_\mathrm{s}}\right)^\gamma  \left( 1 + \frac{r}{r_\mathrm{s}} \right)^{3-\gamma}}; ~~~~\gamma=1.
\end{equation}
The normalization constant is given by the virial mass $M_{\mathrm{vir}}$, which we define using the virial overdensity ($\delta_{\mathrm{vir}})$ \cite{bryan_1998}:
\begin{equation}\label{eq:halomass}
M_\mathrm{vir} = \frac{4\pi}{3}~r_\mathrm{vir}^3~\delta_\mathrm{vir}~\rho_c,
\end{equation}
where $\rho_c$ is the critical density of the Universe at the lens redshift and $r_\mathrm{vir}$ is the virial radius.
The scale radius $r_\mathrm{s}= r_\mathrm{vir}/c$ can be derived assuming a mass-concentration relation, which is:
\begin{equation}\label{eq:cvir}
\log(c) = a + b \log(M_\mathrm{vir}/[10^{12}~h^{-1}~\mathrm{M_\odot}]),
\end{equation}
where $a$ and $b$ are the best fit values for the $z=1$ bin from \cite{dutton_maccio_2014}.

In summary, the parameters describing the mass distribution of the lens are: the halo mass $M_{\mathrm{vir}}$, the SPS-based stellar mass $M_{\ast}^{\mathrm{Chab}}$, the stellar half-light radius $r_{\mathrm{e}}$, and the true stellar mass $M_{\ast}^{\mathrm{True}}$. In practice, we choose to replace the latter with the IMF mismatch parameter \alphaimf\ defined in Eq.~\ref{eq:alphaIMF}.

Following \cite{sonnenfeld_2024}, we defined the prior probability for the lens model parameters, ${\rm P}_{\mathrm{SL}}$, as the product between the distribution in galaxy properties before lensing, ${\rm P}_{\mathrm{g}}$, multiplied by the lens selection probability ${\rm P}_{\mathrm{sel}}$:
\begin{equation}\label{eq:P_SL}
\pr_\mathrm{SL}(M_\mathrm{vir}, M_\ast^\mathrm{Chab}, r_\mathrm{e}, \alpha_{\mathrm{IMF}}) = \pr_\mathrm{g}(M_\mathrm{vir}, M_\ast^\mathrm{Chab}, r_\mathrm{e}, \alpha_{\mathrm{IMF}}) \times \pr_{\mathrm{sel}}(M_\mathrm{vir}, M_\ast^\mathrm{True}, r_\mathrm{e},\alpha_{\mathrm{IMF}}).
\end{equation}
The goal is to determine $\pr_{\mathrm{g}}$.

We modeled $\pr_\mathrm{g}$ as the following product:
\begin{equation}\label{eq:P_g}
\pr_\mathrm{g}(M_\mathrm{vir}, M_\ast^\mathrm{Chab}, r_\mathrm{e}, \alpha_{\mathrm{IMF}}) = \pr(M_\mathrm{vir}) ~\pr(M_\ast^\mathrm{Chab}|M_\mathrm{vir})~\pr(r_\mathrm{e}|M_\ast^\mathrm{Chab}) \pr(\alpha_{\mathrm{IMF}}),
\end{equation}
where $\pr(M_\mathrm{vir})$ is the halo mass function, $\pr(M_\ast^\mathrm{Chab}|M_\mathrm{vir})$ is the stellar-to-halo mass relation, $\pr(r_\mathrm{e}|M_\ast^\mathrm{Chab})$ is the mass-size relation, and $\pr(\alpha_{\mathrm{IMF}})$ is the distribution in IMF mismatch parameter, which we assumed to be independent of galaxy properties.
We computed the halo mass function $P(M_\mathrm{vir})$ with the {\sc{Colossus}} code \cite{diemer_2018} using the \cite{despali_2016} model, sampling a wide uniform distribution $11.5<\log(M_\mathrm{vir}/M_\odot)<14.5$. 
For the stellar-to-halo mass prior we assumed a Gaussian distribution in $\log{M_{\ast}^{\mathrm{Chab}}}$, that is $\mathcal{N}(\log{M_\ast^{\mathrm{Chab}}};\mu_\mathrm{SHMR}, \sigma_\mathrm{SHMR})$. 
The value $\mu_\mathrm{SHMR}(M_\mathrm{vir})$ is the average $\log{M_\ast^{\mathrm{Chab}}}$ for a given halo mass, and $\sigma_\mathrm{SHMR}$ is the scatter around the mean. For the relation we assumed the functional form presented by \cite{shuntov_2022}, with the best fit values and scatter reported for the redshift bin $0.8 < z < 1.1$. Similarly, for the stellar mass-size relation prior we assumed a Gaussian distribution in $\log{r_{\mathrm{e}}}$ at fixed $M_\ast^{\mathrm{Chab}}$, that is $\mathcal{N}(\log{r_{\mathrm{e}}};\mu_\mathrm{MSR}(M_\ast^\mathrm{Chab}), \sigma_\mathrm{MSR})$. The values for $\mu_\mathrm{MSR}(M_\ast^\mathrm{Chab})$ and $\sigma_\mathrm{MSR}$ are obtained from the functional form presented by \cite{van_der_wel_2014}, averaging the best fit values over the $z=0.75$ and $z=1.25$ bins.
Finally, we assumed a Dirac delta function for the distribution in IMF mismatch parameters:
\begin{equation}\label{eq:delta}
\pr(\alpha_{\mathrm{IMF}}) = \delta(\alpha_{\mathrm{IMF}} - \bar{\alpha}_{\mathrm{IMF}}).
\end{equation}
Where $\bar{\alpha}_{\mathrm{IMF}}$ is the general IMF mismatch parameter of galaxies. 

We wrote the selection probability term as the product between the lensing cross-section, $\sigma_{\mathrm{SL}}$, and a GMP-related selection, $\pr_{\mathrm{GMP}}$:
\begin{align} \label{eq:P_sel}
& \pr_{\mathrm{sel}}(M_\mathrm{vir}, M_\ast^\mathrm{Chab}, r_\mathrm{e},\alpha_{\mathrm{IMF}}) \propto \sigma_\mathrm{SL}(M_\mathrm{vir}, M_\ast^\mathrm{Chab}, r_\mathrm{e}, \alpha_{\mathrm{IMF}}) \times \nonumber \\
& \pr_\mathrm{GMP}(M_\mathrm{vir}, M_\ast^\mathrm{Chab}, r_\mathrm{e}, \alpha_{\mathrm{IMF}}).
\end{align}
The lensing cross-section is the source-plane area where a source needs to lie in order to produce two or more detectable images. Following \cite{sonnenfeld_2024}, we computed it for a reference source with intrinsic flux equal to the photometric detection limit, under the assumption of circular symmetry for the lens mass distribution. The latter assumption is justified by the fact that the lensing cross-section is a weak function of the ellipticity of the lens \cite{sonnenfeld_2024}.
The term $\pr_{\mathrm{GMP}}$ describes the probability of a strong lens being discovered with the GMP method, given that it is detected photometrically.
\cite{mannucci_2022} demonstrated that it is well-described by a Heaviside function of the separation of the multiple quasar images (i.e., $\theta_{\mathrm{Ein}}$) with a cut-off at 0.15", so we write it as
\begin{equation}\label{eq:GMPselecfunc}
\pr_{\mathrm{GMP}} \equiv \mathcal{H}(\theta_{\mathrm{Ein}}(M_\mathrm{vir}, M_\ast^{\mathrm{Chab}}, r_\mathrm{e},\alpha_{\mathrm{IMF}}) - 0.15^{"}).
\end{equation}

Accordingly to Bayes theorem, the posterior probability of the parameters describing our lens is defined as:
\begin{align}
& \pr(M_\mathrm{vir}, M_\ast^\mathrm{Chab}, r_{\mathrm{e}}, \alpha_\mathrm{IMF}| M_{\ast,\mathrm{obs}}^\mathrm{Chab}, r_\mathrm{Ein}, r_\mathrm{e}^{(\mathrm{obs})}) \propto 
\pr_{\mathrm{SL}}(M_\mathrm{vir}, M_\ast^\mathrm{Chab}, r_{\mathrm{e}}, \alpha_\mathrm{IMF})\times \nonumber \\
& \pr(M_{\ast,\mathrm{obs}}^\mathrm{Chab}, r_\mathrm{Ein}, r_\mathrm{e}^{(\mathrm{obs})}|M_\mathrm{vir}, M_\ast^\mathrm{Chab}, r_{\mathrm{e}}, \alpha_\mathrm{IMF})\label{eq:bayes},
\end{align}
where the prior probability is the distribution $\pr_{\mathrm{SL}}$ introduced in Eq.~\ref{eq:P_SL}, and $M_{\ast,\mathrm{obs}}^\mathrm{Chab}$ and $r_{\mathrm{e}}^{(\mathrm{obs})}$ are the observed stellar mass and half-light radius, respectively.

The likelihood can be split into a factor that depends on the observed stellar mass, a factor that depends on the Einstein radius, and a term that depends on the observed half-light radius, as follows:
\begin{align}
& \pr(M_{\ast,\mathrm{obs}}^\mathrm{Chab}, r_\mathrm{Ein}, r_\mathrm{e}^{(\mathrm{obs})}|M_\mathrm{vir}, M_\ast^\mathrm{Chab}, r_{\mathrm{e}}, \alpha_\mathrm{IMF}) =  \pr(M_{\ast,\mathrm{obs}}^\mathrm{Chab}|M_\ast^\mathrm{Chab}) \times \nonumber \\
& \pr(r_\mathrm{Ein}|M_\mathrm{vir}, M_\ast^\mathrm{Chab},  r_{\mathrm{e}}, \alpha_\mathrm{IMF}) \pr(r_\mathrm{e}^{(\mathrm{obs})}|r_\mathrm{e}).
\end{align}

We assumed a Gaussian distribution for the stellar mass term, 
\begin{equation}
\pr(M_{\ast,\mathrm{obs}}^\mathrm{Chab}|M_\ast^\mathrm{Chab}) = \mathcal{N}(M_{\ast}^\mathrm{Chab};M_{\ast,\mathrm{obs}}^\mathrm{Chab}, \sigma_\mathrm{M}),
\label{eq:like_Mstar}
\end{equation}
where $\sigma_\mathrm{M}$ is the uncertainty on $M_{\ast,\mathrm{obs}}^\mathrm{Chab}$.
Furthermore, we assumed a Dirac delta function for $\pr(r_\mathrm{e}^{(\mathrm{obs})}|r_\mathrm{e})$, given that the uncertainty on the half-light radius is very small.

We are interested in the value of the IMF mismatch parameter of the galaxy population, $\bar{\alpha}_{\mathrm{IMF}}$. So, we marginalized Eq.~\ref{eq:bayes} over the specific value of the IMF mismatch parameter of our lens, \alphaimf. In practice, this integral is trivial, owing to our modelling choice of Eq.~\ref{eq:delta}. From here onward, we simply set $\alpha_{\mathrm{IMF}} = \bar{\alpha}_{\mathrm{IMF}}$, for the sake of simplicity of notation. We assumed a uniform prior on $\log{\alpha_{\mathrm{IMF}}}$ in the range $\mathcal{U}(-0.1, 0.3)$, encompassing reasonable IMFs values (see Figure \ref{fig:alphaimf}).

We sampled from the posterior probability with a Markov chain Monte Carlo (MCMC), using the {\sc{emcee}} Python package \cite{foreman_2013}.
At each step of the MCMC, we ensured that the prior probability $\pr_{\mathrm{SL}}$ be properly normalised, that is:
\begin{equation}\label{eq:pslnorm}
\int \pr_\mathrm{SL}(M_\mathrm{vir}, M_\ast^\mathrm{Chab}, r_\mathrm{e},\alpha_{\mathrm{IMF}}) ~\mathrm{d}M_\mathrm{vir} ~\mathrm{d}M_\ast^\mathrm{Chab} ~\mathrm{d} r_\mathrm{e} ~\mathrm{d}\alpha_{\mathrm{IMF}} = 1.
\end{equation}
We computed the integral above with Monte Carlo integration and importance sampling. Results for the Chabrier IMF are shown in Figure \ref{fig:MCMC}, where the errors are reported at the 99.7\% level.

 \begin{figure} 
	\centering
	\includegraphics[width=0.9\textwidth]{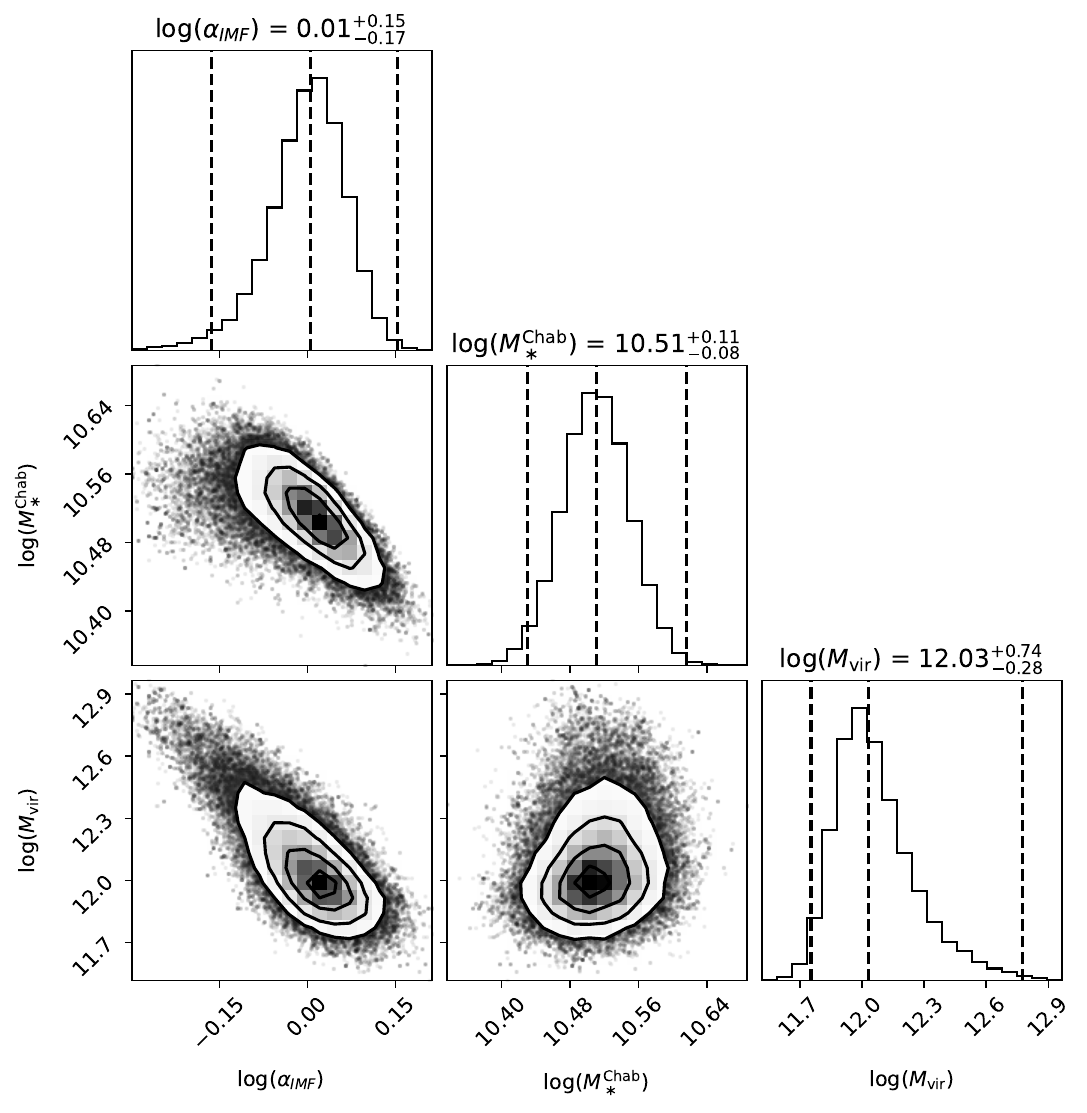} 
	\caption{Corner plot of the posterior probabilities of the MCMC fitting results.. The reported values and uncertainties are the median and the 99.7\% boundaries of the distributions, respectively, that are also marked by the central and external dashed lines in the histograms.
        }
	\label{fig:MCMC} 
\end{figure}

\subsection{The IMF mismatch parameter in absence of dark matter}
\label{subsec:IMF_infer_noDM}

In this exercise there is not a DM halo by construction, so no DM profile or priors given by halo mass function and halo's scale relations are taken into account. The only assumption is given by the best-fit stellar mass profile obtained in Subsection \ref{subsec:light_fit}. The measurable is \Mein~ that, in absence of DM, is assumed to be the total stellar mass within \rein. By defining the following equation:

\begin{equation}\label{eq:alphaimf_noDM}
M_{\ast,\mathrm{Ein}}^\mathrm{True} = \alpha_\mathrm{IMF} \times M_{\ast}^\mathrm{Chab} \times f_\mathrm{Ein} ~, 
\end{equation}
we aim at sampling the posterior probability of \alphaimf. $f_\mathrm{Ein}$ is the fraction of stellar mass contained in \rein, given the de Vaucouleurs profile. As a prior for \alphaimf~ we assumed a flat distribution similarly to what is shown in Subsection \ref{subsec:IMF_infer}. For $M_{\ast}^\mathrm{Chab}$ we assumed a Gaussian likelihood centered on the observed value (see Eq. \ref{eq:like_Mstar}), while for $M_{\ast,\mathrm{Ein}}^\mathrm{True}$ we assumed a Gaussian likelihood centered on \Mein~ (with standard deviation equal to its 1$\sigma$ error). We performed an MCMC sampling of the \alphaimf~ posterior probability using {\sc{emcee}}, which in logarithmic space returns $\log$(\alphaimf) $= 0.20 ^{+0.03}_{-0.03}$ at 68\% significance level.

\subsection{Dark matter fraction}
\label{subsec:DM_frac}

We evaluated the DM fraction at a given radius $r$, defined as:
\begin{equation}\label{eq:f_DM}
f_\mathrm{DM,r}=\frac{M_\mathrm{DM,r}}{M_\mathrm{tot,r}}
\end{equation}
where $M_\mathrm{tot,r} = M_\mathrm{DM,r} + M_\mathrm{\ast,r}$. We note that in general $0 \leq f_\mathrm{DM,r_{Ein}} \leq 1$ by definition. However, at \rein~ we have a physical constrain on the \emph{projected} total mass: $M_\mathrm{tot,r_{Ein}} = M_\mathrm{Ein}$.
We calculated \Meinstar as the integral from $r=0$ to \rein\ of the circularized $n=4$ \sersic~ light profile used to fit the lens image, converted to a mass profile for different IMFs. To measure values and statistical uncertainties of \Meinstar~ and \fDM, we sampled the Bayesian posterior probabilities of the involved quantities (\reff~ and \rein) from the image and lensing fitting described above. The measured values (median of the probability distribution) are reported in Table \ref{tab:DM_frac}. For each one, the statistical uncertainties are reported at the 99.7\% confidence. However, the measurement of the total stellar mass is largely dominated by the systematic uncertainty given by the conservative assumption that we made on the flux calibration zero-point, which affects the total flux of the source. Thus, we decided to report this error separately.

\begin{table}
	\centering
	\caption{Projected Stellar masses and dark matter fractions  within \rein~ for different IMF assumptions.
}
	\label{tab:DM_frac} 

\begin{tabular}{crc}
\hline
\multicolumn{3}{c}{Chabrier IMF}\\
\hline
\Meinstar & $1.29~ (^{+0.07}_{-0.07})_\mathrm{stat}~(^{+0.17}_{-0.08})_\mathrm{sist} \times 10^{10}$ & $M_\odot$ \\
\fDM & $0.36~ (^{+0.08}_{-0.08})_\mathrm{stat}~(^{+0.04}_{-0.08})_\mathrm{sist}$ & \\
\hline
\multicolumn{3}{c}{Salpeter IMF}\\
\hline
\Meinstar & $2.34~ (^{+0.07}_{-0.07})_\mathrm{stat}~(^{+0.16}_{-0.17})_\mathrm{sist} \times 10^{10}$ & $M_\odot$ \\
\fDM & $-0.14~ (^{+0.14}_{-0.15})_\mathrm{stat}~(^{+0.12}_{-0.08})_\mathrm{sist}$ & \\
\hline
\end{tabular}
\end{table}

The deprojected DM fraction at \reff~ (\fDMdep) is calculated by normalizing a general NFW profile (\ref{eq:nfw}), to the projected DM mass at \rein~ calculated as \Mein $-$ \Meinstar. The scale radius is calculated accordingly to Eq. \ref{eq:cvir}. the \fDMdep~ is then calculated as:
\begin{equation}\label{eq:f_DM_3D}
f_\mathrm{DM, 3D,r_e}=\frac{M_\mathrm{DM, 3D,r_e}}{M_\mathrm{DM, 3D,r_e}+M_\mathrm{\ast, 3D,r_e}},
\end{equation}
where $M_\mathrm{DM, 3D,r_e}$ and $M_\mathrm{\ast, 3D,r_e}$ are the DM and stellar deprojected masses, calculated assuming spherical symmetry. For the stellar component a de Vaucouleurs profile is used, as found from the light fitting. For profiles other than a NFW ($\gamma = 1$) the contraction is adjusted by varying the profile slope. As shown by \cite{sonnenfeld_2025}, a maximum contraction allowed by hydrodynamical simulations corresponds to $\gamma\sim1.7$. 

\subsection{Constraints on the galaxy age}
\label{subsec:gal_age}

The galaxy formation redshift assumed by \cite{longhetti_2009} to calculate the M/L conversion factor that we used in this work is $z_\mathrm{f} =4$. 
While for older galaxies ($z_\mathrm{f} >4$), increasing the observed stellar masses would lead to a further deviation from the true value, for a younger source the derived stellar mass with a Salpeter IMF can be equal to that derived with a Chabrier IMF and a higher $z_\mathrm{f}$. By using the age-depended M/L conversion factor reported by \cite{longhetti_2009}, we restricted the minimum galaxy age at which the Salpeter IMF is rejected at the 99.7\% level at $\sim$3.3 Gyr ($z_\mathrm{f}>2.9$).
If the galaxy is $\sim$3 Gyr old ($z_\mathrm{f}\sim2.5$), instead, the Salpeter IMF is rejected at the 90\% level. 
For a lower formation redshift, a Salpeter IMF cannot be ruled out with high significance: given that the timescale for stellar mass assembly is expected to be short ($<1$ Gyr), the galaxy could have formed its stellar mass within a few hundred million years.

With the current data it is impossible to obtain an independent direct measurement of the stellar population age. However, we built our Bayesian framework to infer \alphaimf~ on the basis of several scaling relations, which provided us with the virial mass of the DM halo: $\log(M_\mathrm{vir})= 12.03^{+0.78}_{-0.28}$ at 99.7\% confidence level. Thus, we can trace back the evolution of the halo to find its formation redshift $z_\mathrm{f,~halo}$ and thus a lower limit to the galaxy age, since hydrodynamical simulations (implementing stellar and AGN feedback) show that the star formation starts right after the halo collapse and shuts down rapidly after black-hole feedback activates, in a time span of less than 1 Gyr \cite{donnari_2019}. The mass accretion history (MAH) of DM halos has been extensively studied in DM-only (DMO) and hydrodynamical simulation. A definition of $z_\mathrm{f,~halo}$ commonly used in simulations to select already relaxed systems, referred to as the virial criterion, is the redshift at which the mass of the halo progenitor equals the mass enclosed in the scale radius of a given DM profile at the observing redshift \cite{ludlow_2013}. Based on extended Press–Schechter theory and DMO numerical simulations, \cite{correa_2015} investigated the dependence of several properties of a NFW halo observed at a given redshift -- most importantly the halo mass and concentration -- on $z_\mathrm{f,~halo}$. Importantly, they selected relaxed systems on the basis of friend-of-friend halo separation threshold in their numerical simulations, showing that the MAH-concentration relation, essentially linked by $z_\mathrm{f,~halo}$, is almost independent from the definition of relaxed halos for galactic-scale and low-redshift ($z<3$) observed objects, while including non-relaxed systems may have some impact on high-$z$ proto-clusters. Exploiting these relations, for the redshift, virial mass and concentration of our halo, we find $z_\mathrm{f,~halo} = 3.75_{-0.70}^{+0.25}$, where the uncertainties are drawn from the extrema of the relations scatter. This finding is fully in agreement with the galaxy $z_\mathrm{f} =4$ assumed by \cite{longhetti_2009}, and with the minimal galaxy age that we require to reject the Salpeter IMF at 3$\sigma$. In addition, \cite{Beltz-Mohrmann_2021} investigated the impact on the DMO-derived mass-concentration relation (and thus on $z_\mathrm{f,~halo}$) when baryonic physics is considered, exploiting a suite of state-of-art hydrodynamical simulations (Illustris, TNG100 and EAGLE), concluding that the mass-concentration slopes in hydrodynamical and DMO simulations are consistent. Finally, we can provide an additional estimate of the halo mass from an independent spectral measurement, that is the EW ${\sim}$2.2 $\mathrm{\AA}$ of the Mg~II]$\lambda2796\mathrm{\AA}$ absorption line. \cite{das_2025} investigated the relation between this EW and the DM halo virial mass, using our same halo mass function and a consistent stellar-to-halo mass relation \cite{girelli_2020}, in a wide redshift range $0.4 < z < 1$. For $1.5\mathrm{\AA <}$ EW $ <2.5 \mathrm{\AA}$, they find a mean DM halo mass distribution peaking at $\log(M_\mathrm{vir}) = 11.5$ with a 0.3 scatter, in excellent agreement with our estimate considering the large EW and redshift bins. We remark that, as noted by \cite{correa_2015}, a smaller halo mass would imply an earlier $z_\mathrm{f,~halo}$, when the Universe was denser and halos more concentrated.

\subsection{Cosmology}
\label{subsec:cosmology}

Throughout the work we adopt a concordance $\Lambda$CDM cosmology with $\mathrm{H_0} = 70~\mathrm{km~s^{-1}~Mpc^{-1}}$, $\Omega_{\mathrm{M}} = 0.3$, and $\Omega_{\mathrm{\Lambda}} = 0.7$, in agreement with the \textit{Planck 2018} results \cite{PLANCK_2018}.

\bmhead{Acknowledgements}
Based on observations collected at the European Organisation for Astronomical Research in the Southern Hemisphere under ESO programmes  111.24QJ and 112.25GT (PI: Mannucci). Based on observations made with the Italian Telescopio Nazionale Galileo (TNG) operated on the island of La Palma by the Fundación Galileo Galilei of the INAF (Istituto Nazionale di Astrofisica) at the Spanish Observatorio del Roque de los Muchachos of the Instituto de Astrofisica de Canarias.
QD acknowledges the ``Progetto Giovani Astronomi'',  aiming at promoting scientific activities for high-school students, by offering them the opportunity to observe selected targets at the TNG under professional astronomer supervision. Chiara Izzo, Emanuele Morone, Annalisa Vitelli, Italo Grasso, Fabio Pascale, Leonardo Karol Iaquinto are the students who performed the observations on site. 
QD acknowledges Dr. Laura Scholz-Díaz for the valuable discussion regarding the comparison with the simulations.

\section*{Declarations}

\begin{itemize}
\item Funding: QD, FM and PS acknowledge financial support from the INAF Bando Ricerca Fondamentale INAF 2022 Large Grant: ``Dual and binary supermassive black holes in the multi-messenger era: from galaxy mergers to gravitational waves'' and from the INAF Bando Ricerca Fondamentale INAF 2024 Large Grant: ``The Quest for dual and binary massive black holes in the gravitational wave era''.
C.~S. acknowledges financial support from INAF $-$ Ricerca Fondamentale 2024 (Ob.Fu. 1.05.24.07.04) and by the Italian Ministry of University and Research (grant FIS-2023-01611, CUP C53C25000300001). 
JWN is supported by an STFC/UKRI Ernest Rutherford Fellowship, Project Reference: ST/X003086/1. 
GV and SC acknowledge support by European Union’s HE ERC Starting Grant No. 101040227 - WINGS.
GC and EB acknowledge the support of the INAF Large Grant 2022 "The metal circle: a new sharp view of the baryon cycle up to Cosmic Dawn with the latest generation IFU facilities". 
EB acknowledges INAF funding through the “Ricerca Fondamentale 2024” program (mini-grant 1.05.24.07.01). 
MP acknowledges support through the project PID2021-127718NB-I00 and the grant RYC2023-044853-I, funded by  MCIN/AEI/10.13039/501100011033 and El Fondo Social Europeo Plus.
IL acknowledges support from PRIN-MUR project "PROMETEUS"  financed by the European Union -  Next Generation EU, Mission 4 Component 1 CUP B53D23004750006.
\item Conflict of interest/Competing interests: None 
\item Data availability: We report the ID of the ERIS observing program regarding the data used in this work, publicly available at the ERIS ESO archive:
SPIFFIER Data: program ID = 111.24QJ.001; 
NIX Data: GTO program ID: 112.25GT.001.
The TNG data are not available at the TNG archive, as they are part of the ``Giovani Astronomi'' project. Data are available at the Zenodo public repository (Dataset DOI: 10.5281/zenodo.16213717).
\item Code availability: The software used for the data analysis are all public and retrievable through the references provided in the text.  
\item Author contribution: QD led the conceptualization of the work, data reduction and analysis, investigation, and writing the original draft. FM contributed to the conceptualization, analysis, draft revision and funding acquisition. AS contributed to conceptualization, analysis, draft writing and revision. MS contributed to data reduction and analysis, and draft revision. JN contributed to data analysis and draft revision. CS, SZ, AM, PR, AG, ED contributed to conceptualization, investigation, data interpretation and draft revision. CM, GA contributed to data analysis and draft revision. FB, EB, CB, SC, EC, AC, MC, CC, AC, GC, AD, AF, MG, IL, BM, EN, MP, RR, PS, PS, GT, GV, LU, CV, MZ contributed to data interpretation and draft revision. GA, ED, EP, VT, AP contributed to the data acquisition and observations.

\end{itemize}







\begin{appendices}

\section{Literature comparison}\label{secA1}

\subsection{Sources of uncertainty for different methods}
\label{subsec:uncer_comp}

Many analyses performed over the past two decades have tended to favor a Salpeter IMF for massive ($M_\ast \geq 10^{11}~\mathrm{M_\odot}$) ETGs at low redshift. The majority of these results, starting from the pioneering studies presented by \cite{TK_2004, koopmans_2006, treu_2010a}, are based on joint LD analyses, particularly of SLACS systems, which is by far the most studied sample of lensing galaxies \cite{treu_2010a}. The same conclusion has been reached by dynamics-only analysis studies of similarly massive samples in the local Universe (such as in the case of ATLAS$^{\rm 3D}$ survey, \cite{cappellari_2012,cappellari_2013}). 

Interpreting the stellar kinematics of galaxies is a challenging task, as it requires reconstructing their three-dimensional mass distribution and the underlying stellar orbital structure -- quantities that cannot be observed directly. The studies based on the SLACS sample, in particular, were mostly based on single-aperture stellar velocity dispersion ($\sigma_\ast$) measured from Sloan Digital Sky Survey (SDSS). The lack of spatially resolved spectroscopic data implied the adoption of simplifying assumptions, most commonly spherical symmetry, and in some cases even orbital isotropy \cite{auger_2010b, sonnenfeld_2015}. However, spatially resolved follow-ups of a SLACS subsample showed that these assumptions lead to an incorrect density slope determination \cite{barnabe_2011}, meaning that they are not sufficiently accurate for the majority of galaxies in the sample. In addition, a recent study of the SLACS sample demonstrated that LD studies are also biased by selection. This is primarily made on the galaxy dynamics, favoring the inclusion of galaxies with high $\sigma_\ast$; as a result, LD samples are not representative of the overall ETG population \cite{sonnenfeld_2024}. Since both LD and dynamics-only studies showed a positive correlation between the M/L and $\sigma_\ast$ \cite{posacki_2015,mendel_2020}, this result implies that LD studies based on dynamically-selected samples (such as SLACS) tend to skew the average value towards heavier IMFs.
Another fundamental source of uncertainty in IMF studies is the sensitivity of different techniques, and even different samples analyzed with the same technique, to different physical scales. 
LD samples, especially those with only single-aperture stellar velocity dispersion available, are particularly affected by this uncertainty, as lensing and dynamics probe different scales: while lensing measures the mass enclosed within \rein, dynamics is more sensible to the mass within \reff~ \cite{shajib_2021, sonnenfeld_2024}. 
In the presence of M/L radial gradients (as predicted by the two-phase evolution scenario), this factor can lead, or contribute, to conflicting results obtained even when the same techniques and assumptions are applied to different samples, if their Einstein-to-effective radius ratio $R$ is significantly different. For example, the lensing-only analysis of the Hyper SuprimeCam (SuGOHI) sample, featuring a large $R\gg1$, showed that the sample is well described by a bottom-light IMF \cite{sonnenfeld_2019} when a NFW profile is assumed for the DM. 
In contrast, a recent lensing-only re-analysis of the SLACS sample (for which $R\sim1$) concluded that a Salpeter IMF is favored, when the same DM profile is assumed \cite{sonnenfeld_2025}. This aspect further complicates the determination of the IMF in the central part of ETGs, since integrated dynamics is always sensible to large scales and lensing object with small $R$ are scarce in the nearby Universe and unknown at high $z\gtrsim0.5$. All the aforementioned LD and dynamics studies (involving the SLACS and ATLAS$^{\rm 3D}$ surveys) were carried under the assumption of a constant M/L, regardless the radius sampled by the observables, contributing to the inhomogeneity of the results.

Lensing-only studies and selection have the advantage of avoiding dynamics-based selection bias and, in absence of spatially-resolved stellar kinematics, of not being subject to the uncertainties implied by the oversimplified assumptions on the stellar dynamics. 
The main source of uncertainty in lensing-only is the assumption of the DM profile which, in any case, also affects LD and dynamics studies (where the DM is often assumed negligible or at fixed fraction at any radii, see \cite{cappellari_2013, tortora_2018, forrest_2022}). 
Most of IMF studies that model the DM, regardless the used technique, assume a NFW profile, as it is generally considered valid for galactic-scale halos \cite{TK_2004, treu_2010a, sonnenfeld_2015, shajib_2021, mendel_2020}. 
However, both N-body and hydrodynamical simulations consistently predict halo contraction, counteracted by star formation feedback \cite{cautun_2020,schaller_2015}.
A recent free-halo lensing-only reanalysis of the SLACS sample showed that a IMF-halo contraction degeneracy is present, comparable to the Chabrier–Salpeter offset \cite{sonnenfeld_2025}. 
However, in lensing-only studies the DM fraction at a given radius can also be inferred by hydrodynamical simulations, which account for halo contraction and baryonic feedback; moreover, extreme assumptions can be made on the DM (such as removing it), to derive stringent constraints to the IMF. 
In addition, depending on the size of \rein, lensing can sample the baryon-dominated inner part of galaxies, where the degeneracy with the DM is less critical and M/L gradients negligible. For the SNELLS survey analysis, prescriptions obtained from the Evolution and Assembly of Galaxies and Their Environments (EAGLE, \cite{schaye_2015}) hydrodynamical simulation have been used to infer the DM fraction within \rein, and infer the bulge IMF \cite{smith_2015}. 
In contrast with previous LD studies on comparable mass and $\sigma_\ast$ (but on average different $R$) ETGs, a bottom‐light (Kroupa‐like) IMF is found if the galaxies are old. 
The sample has been subsequently re-analyzed by \cite{newman_2017} with several techniques, including LD and dynamics-only, finding that when dynamics is included the IMF is systematically heavier than that obtained with the lensing-only analysis. 
In addition, the SPS fitting of the spectra provides contrasting results, scattering between heavier-than-Salpeter to Kroupa IMF depending on the assumed model and analyzed object. 
This aspect highlights another important source of uncertainty in IMF determination studies, consisting of the several (and in general different from work to work) assumptions made to model the stellar population, with special regard to its star formation history and age. 
As for the discrepancy between the lensing and dynamics, given all the aforementioned source of uncertainties for the dynamics modeling with respect to the lensing-only, for the SNELLS re-analysis the authors conclude that the lensing masses are more precise and less sensitive to the modeling assumptions than the dynamical masses \cite{newman_2017}. 
This is especially true considering that, by removing the DM (i.e. the main source of uncertainty in lensing), they still find an IMF significantly lighter than Salpeter. 
In summary, considering the limitations of the stellar kinematic constraints and analysis, the lensing constraints are in general considered more robust \cite{newman_2017,shajib_2024}. 
Other kinematic studies have inferred IMF normalizations using molecular gas dynamics \cite{davis_2017}, which in some systems require light IMFs to avoid stellar mass estimates exceeding dynamical mass, while others imply heavier IMFs, highlighting the heterogeneity of results with these methods. All the discussed possible sources of uncertainties can sum up to contribute to the observed bias between the lensing and dynamics, and a dominant factor (if any) has not yet been identified.

\subsection{Comparison with the velocity dispersion of LD galaxies}
\label{subsec:LD_comp}

We do not have information about the central stellar velocity dispersion of J1453g. However, given the assumed SIE model, the relation between the Einstein radius and the SIE velocity dispersion $\sigma_\mathrm{SIE}$ is analytical \cite{ferrami_2025}:
\begin{equation}
r_\mathrm{Ein} (\sigma_\mathrm{SIE}, z_\mathrm{l}, z_\mathrm{s}) = 4\pi \frac{D_\mathrm{ls}}{D_\mathrm{s}} \left( \frac{\sigma_\mathrm{SIE}}{c} \right)^2.
    \label{eq:v_SIE} 
\end{equation}
Thus, we can derive it for our object and compare the \alphaimf~ value found by our Bayesian inference with that expected from LD studies, in particular for massive ETGs at $z{\sim}0.3$ from the SLACS survey that provides $\sigma_\mathrm{SIE}$ measurements. We derive $\sigma_\mathrm{SIE} = 127 \pm 2$ km/s. Extrapolating the results of \cite{treu_2010a} for such a $\sigma_\mathrm{SIE}$ we found a log(\alphaimf) $\sim 0.02$.

\section{Additional sources of uncertainty}\label{secA2}

\subsection{Caveats on the dark matter fraction calculation}
\label{subsec:DM_caveats}

While it is true that the projected DM mass at \rein\ can be simply derived as the difference between the total mass and the stellar mass, without making assumption on the DM profile, the mass \Mein\ is derived from the lens reconstruction method that assumes an overall total mass profile. Hence, by adopting a stellar profile for the mass component and a total mass profile, we are implicitly making assumptions about the DM distribution. In our analysis we assume a SIE for the total mass. However, although we fix the slope in the fit to better constrain \rein, leaving it free still yields results consistent with SIE within 1$\sigma$. Thus, our working hypothesis is that the ``bulge–halo conspiracy'' is valid, a conclusion that has been repeatedly confirmed by lensing studies across a wide range of masses and redshifts \cite{TK_2004, koopmans_2006,auger_2010b,sonnenfeld_2012,sonnenfeld_2013_III,shajib_2021,shajib_2024,tan_2024}. At any rate, we note that the only lensing-derived quantity that we use in the straightforward DM mass calculation is \Mein, which is nearly independent on the used mass model as it depends predominantly on the monopole of the lens potential \cite{kochanek_1991, treu_2004}.

\subsection{Implication of the stellar size uncertainty}
\label{subsec:stellar_size_unc}

We used a de Vaucouleurs profile for the light fitting of our object in $Ks$ band. This is consistent with the fitting of almost all the samples we compared with in terms of IMF normalizations (SLACS, SL2S, SNELLS; \cite{smith_2015,sonnenfeld_2015}). The only exception is the SuGOHI sample \cite{sonnenfeld_2019}, for which a \sersic~ $n = 1$ is assumed (i.e., late-type galaxy typical value). We note that different assumptions on the stellar profiles do not change our conclusions: we tested different \sersic~ profiles models to assess the variation in the \fDM, finding that for a extremely high $n = 10$ the \fDM~ only decreases by 4\%, while decreasing the $n$ would increase the \fDM, requiring an even lighter IMF than a Chabrier. In addition, the analysis of the SL2S sample performed by \cite{sonnenfeld_2013_III} shows how the galaxy \reff~ for profiles other than \sersic~ (i.e., Hernquist) varies at most 10\%. The uncertainty on \reff~ is dominated by the photometric band used; in the same analysis of the SL2S sample, performed in nearly all bands from $u$ to $K$, \cite{sonnenfeld_2013_III} showed that the \reff~ scatter across this wide range of wavelengths is ${\sim}30$\%. Even assuming the extreme lower boundary of \reff~ due to this uncertainty, the \fDM~ at \rein~ for a Chabrier IMF would decrease at most to ${\sim}$22\%, while assuming a Salpeter IMF would return a very unphysical \fDM ${\sim}-$40\%. By assuming the extreme upper boundary, instead, we obtain \fDM ${\sim}$47\% and \fDM ${\sim}$3\% for a Chabrier IMF and a Salpeter IMF, respectively. Also in this case, the value obtained for the Salpeter IMF is strongly disfavored by the comparison with all the lensing, LD, and dynamic results for massive galaxies at any redshift, and by hydrodynamical simulation predictions. In summary, our conclusions on the bottom-light IMF of J1453g is not affected by the uncertainty on the galaxy size and stellar profile.





\end{appendices}


\bibliography{lens}

\vspace{30pt}
\noindent \textbf{Affiliations}
\begin{enumerate}
\item INAF -- Osservatorio Astrofisico di Arcetri, Via Largo E. Fermi 5, Firenze, 50125, Italy
\item Department of Astronomy, Shanghai Jiao Tong University, Shanghai, 200240, China
\item University of Trento, Via Sommarive 14, Trento, 38123, Italy
\item Dipartimento di Fisica e Astronomia, Università di Firenze, Via G. Sansone 1, Sesto Fiorentino, 50019, Italy
\item School of Mathematics, Statistics and Physics, Newcastle University, Newcastle-upon-Tyne, NE1 7RU, UK
\item INAF -- Istituto di Radioastronomia, Via Piero Gobetti 101, Bologna, 40129, Italy
\item Dipartimento di Fisica e Scienze della Terra, Università di Ferrara, Via G. Saragat 21, Ferrara, 44122, Italy
\item European Southern Observatory, Karl-Schwarzschild Straße 2, D-85748 Garching bei München, Germany
\item INAF -- Fundación Galileo Galilei, Rambla José Ana Fernández Pérez 7, Breña Baja, 38712, Spain
\item Scuola Normale Superiore, Piazza dei Cavalieri 7, Pisa, 56126, Italy
\item Institute of Theoretical Astrophysics, University of Oslo, PO Box 1029, Blindern, Oslo, 0315, Norway
\item Department of Physics and Astronomy, University of California, Los Angeles (UCLA), 430 Portola Plaza, Los Angeles, CA, 90095, USA
\item INAF -- Istituto di Astrofisica e Planetologia Spaziali, Via Fosso del Cavaliere 100, Rome, 00133, Italy
\item INAF -- Osservatorio Astronomico d'Abruzzo, Via Mentore Maggini snc, Teramo, 64100, Italy
\item Centro de Astrobiología, CSIC--INTA, Cra. de Ajalvir Km. 4, Torrejón de Ardoz, Madrid, 28850, Spain
\item INAF -- Osservatorio Astronomico di Brera, Via Brera 28, Milano, 20121, Italy
\item INAF -- Osservatorio Astronomico di Roma, Via Frascati 33, Monte Porzio Catone, 00078, Italy
\item Max-Planck-Institut für Extraterrestrische Physik, Giessenbachstraße 1, Garching, 85748, Germany
\item Dipartimento di Fisica e Astronomia, Università degli Studi di Bologna, Via Gobetti 93/2, Bologna, 40129, Italy
\item Istituto d'Istruzione Superiore Telesi@, Via Caio Ponzio Telesino 26, Telese Terme, 82037, Italy
\end{enumerate}

\end{document}